\documentclass[3p,onecolumn]{elsarticle}

\def\part{\partial}

\usepackage{amssymb}

\journal{New Astronomy Reviews}

\begin{document}

\begin{frontmatter}



\title{Persistent  mysteries of jet engines, formation, propagation,  and particle acceleration: have they been addressed experimentally?}


\author{Eric G. Blackman}

\address{Department of Physics and Astronomy, University of Rochester, Rochester NY, 14627, USA}

\author{Sergey V. Lebedev}

\address{Department of Physics, Imperial College London, South Kensington Campus, London UK, SW7 2AZ}

\begin{abstract}

The  physics of astrophysical jets can be divided into three regimes:  (i) engine and launch  (ii) propagation and collimation,  (iii) dissipation and particle acceleration. Since astrophysical jets comprise a huge range of scales and phenomena, practicality dictates that most studies of jets  intentionally or inadvertently  focus on one of these  regimes, and even therein, one body of work  may be  simply  boundary condition for another.  We first discuss long standing persistent mysteries that pertain the physics of each of these regimes, independent of the method used to study them.  This discussion makes contact with  frontiers of plasma astrophysics more generally.     While observations  theory, and simulations, and  have long been the  main tools of the trade,   what about laboratory experiments?  Jet related experiments have  offered controlled studies of  specific  principles, physical processes, and  benchmarks for numerical and theoretical calculations.   We discuss what has been done to date on these fronts.  Although experiments  have indeed helped us to understand certain   processes, proof of principle concepts, and benchmarked codes,  they have  yet to solved  an astrophysical jet mystery on their own.  
 A challenge is that experimental tools used for jet-related  experiments so far, are  typically not machines originally designed for that purpose, or designed with specific astrophysical mysteries in mind. This presents an opportunity for a different way of thinking about the development of future platforms: start with the astrophysical mystery and build an experiment to address it. 
 \end{abstract}

\begin{keyword}
 jets \sep laboratory astrophysics \sep accretion \sep magnetic fields \sep young stellar objects \sep active galactic nuclei\sep microquasars \sep particle acceleration \sep high energy density physics


\end{keyword}

\end{frontmatter}



\section{Introduction}
\label{introduction}

Astrophysical jets are observed over a broad range of scales but 
 form on scales below which most all observations can resolve.  
 Jets occur  in the  early phases of star formation \cite{Bally+2007}; in the late stages of dying massive stars (supernovae, gamma-ray bursts)  \cite{Woosley+2006,Wang+2008} and pulsars \cite{Kargaltsev+2003,Durant+2013}; 
dying low mass stars (pre-planetary and planetary nebulae) \cite{Morris1987,Sahai+1998,Bujarrabal+2001,Guerrero+2020}; and post-stellar compact objects (microquasars)  \cite{Corbel2011} .  Compact object engines of gamma-ray bursts (GRB), microquasars and active galactic nuclei (AGN) all power relativistic jets.  
 See Refs. \cite{deGouveiaDalPino2005,Beall2016,Romero+2017} for  reviews (and this special volume). 
With the exception of  recent observations  of jet of M87
by  the Event Horizon Telescope (EHT) \citep{Jimenez+2018}, 
there is a dearth of   spatially resolved jet observations on scales below   $R\le 50R_{in}$ for any system.
The fact that the very inner scales
are difficult to probe,  contributes to the persistence of mysteries  in understanding 
the jet engines and their larger scale consequences. 


Conceptually and practically, it is useful  to categorize   jet physics into
the following regimes that describe the progression  of the jet flow:  (i) engine and launch
to (ii) propagation and collimation  and  (iii) dissipation and particle acceleration.
For most all jets of astrophysics, role of magnetic fields  as  
drive belts of energy and angular momentum transfer  from engine to jet, and/or as jet collimators 
 This conclusion often emerges  from kinematic considerations, as   the available radiation pressure is  often incapable of providing the
directed  momentum.  Engines powered by accretion with the needed magnetic fields 
generated or advected therein are the leading paradigms.

Even if the launch is magnetically facilitated at the base
 there are more than one way for  this to occur and we cannot claim to understand jet formation until
 we understand not just what is possible but what is happening in practice.
There is also more agreement that magnetic fields are  dynamically significant 
within   the  launch region ($R\le 50 R_{in}$),  than beyond. 
 If for example, by $\sim 50R_{in}$, a magnetically  
driven  outflow is   collimated  at launch and reaches a 
supermagnetosonic speed, the tangent of the jet opening angle
would simply be the inverse supermagnetosonic Mach number  at that point. The jet would  already be collimated and 
perhaps the need to further magnetically collimate beyond that point is obviated.  However, which  specific magnetically dominated launch mechanism dominates at the base matters because they differ as to how magnetically dominated the flow remains.  This also    determines what kind of particle acceleration mechansims
are viable.  

As will be further discussed, our approach differs from two  previous reviews on laboratory jet experiments \cite{Bellan2018} and pulsed power experiments relevant to astrophysics  \cite{Remington+2006,Lebedev+2019} both in topical coverage but also in approach:  we start with a broad discussion of the astrophysical jet mysteries first and then move to what the experiments have addressed, rather than starting with the experiments and identifying applications.  
In sections 2,3, and 4 we  discuss    persistent long standing questions in jet astrophysics 
in the three respective regimes mentioned above: engine  and launch; propagation and collimation, and dissipation and particle acceleration. In section 5 we  discuss what experiments have taught us so far.
Overall, we find  that experiments have impressively addressed  a number of   basic processes, but the challenging opportunity remains  for future experiments to  directly resolve an astrophysical question.   
We conclude in section 6.






\section{Launch Region Mysteries}

The physics of the launch region includes accretion physics, the  origin of large scale magnetic fields, corona formation and  the   physics of distinct magnetically mediated jet launch mechanisms.
 
 \subsection{In accretion engines, where do large scale magnetic fields come from and what is their role in angular momentum transport?}

Due to the inefficiencies of pure hydrodynamics and radiation, magnetic fields have long been  regarded as the  essential intermediary facilitating accretion and  jet outflow  from accretion engines.
Most  magnetically mediated jet models involve 
Poynting flux driving near the disk, employing a field   of large scale with respect to the disk height.
Traditionally, most jet theory and simulation  designed to test magnetically mediated aspects of the jet
 assume that the field is already of large scale as an initial condition.
 
Accretion disks are essential for most classes of  jetted systems (a counter example
may be thermally driven, common-Envelope in-spiral scenarios).
Transport models of angular momentum in accretion disks  include 
mechanisms that appeal to turbulence or turbulent viscosity, as initially framed in the  Shakura-Sunyaev model \cite{Shakura+1973},
or nonlocal   transport involving large scale fields.  Early jet models were in fact presented
for laminar disks as a way to transport angular momentum nonlocally \cite{Blandford+1982,Konigl1989}, while  mesoscale transport models involving coronal reconnection have also been invoked.
\cite{Lynden-Bell1969,Field+1993}.
Probably some combination of the three is ultimately involved \cite{Blackman+2015}.
 
 In the Shakura-Sunyaev framework, the details of the viscosity are not  part of the model,
 although it was recognized that there must be a turbulent enhancement over  microphysical values
 to shorten accretion time scales.   \cite{Balbus+1991,Balbus+1998} introduced the astrophysics community to the magnetorotational instability discovered by Velikhov \citep{Velikhov1959}, which is now widely perceived to be a ubiquitous source of the needed turbulence for sufficiently ionized regions of disks.   But how exactly the magnetorotational instability contributes to  angular momentum transport in real disks still remains unclear.    Despite  decades of simulations of the MRI, modelers in need of practical equations typically default back to the Shakura-Sunyaev formalism that is decoupled from MRI-specific physics.  To what extent is the transport local vs. large scale and on what does this depend? Theories in common use do not offer predictive power as 
 to what fraction of the transport occurs in the disk, the corona, or in jets, and yet observations conspicuously   show spectral evidence for  three  features e.g. \cite{Gierlinski1999}.
Such spectra for  microquasars  and AGN are typically modeled by combining  a linear combination of  jet, corona, and disk spectra  with the relative weighting tuned to match the observations.  There is also not yet clarity on just how the  MRI  contributes to  local, mesoscale, or nonlocal stresses. 
 
 Although even non-local transport can   be replaced by a Shakura-Sunyaev 
$\alpha$ type formalism,  the difference of local vs. non-local transport manifests   in  predicted energy spectra.  If the transport is local then  the associated dissipation in an optically thick disk produces thermal emission. If the transport is nonlocal and occurs after magnetic buoyant rise of structures into  a low density, low plasma $\beta$ corona, then non-thermal acceleration and radiation are possible. 
And for the largest scale transport via jets, the energy dissipation could occur even farther from the
engine, producing yet a different non-thermal  component.

Most magnetized jet models, whether on not jets dominate the extracted accretion power, 
 invoke large scale ordered fields \cite{Blandford+1982,Konigl1989,Pudritz+2012}. 
 Two possibilities for  the origin of the field can be debated:
(1) Is the large scale  field  generated   by a large scale dynamo
\cite{Pudritz1981,Ruediger1993b,Campbell1999,Campbell2000}, possibly  driven 
by  the MRI itself 
\cite{Brandenburg+1995,Lesur2010,Davis2010,Simon2011,Guan2011,Sorathia2012,Suzuki2014,Shi+2016,Bhat+2016}?  All MRI simulations of sufficiently large scale and resolution show evidence
for large scale dynamo action that correlates with saturation of the transport coefficients \cite{Ebrahimi+2009,Guan2011,Ebrahimi2014,Bhat+2016}.
The MRI produces a spectrum of large and small scale magnetic fields and it could 
be that  the  fraction of the spectrum whose scale is large enough to buoyantly rise to the corona before turbulently diffusing  are able to supply coronal magnetic fields which then relax into large scale jet sustaining fields. \cite{Blackman+2009}.    

(2) Is the large scale field advected?
Whether flux
can be advected by the accretion flow has been a long standing question \cite{Lubow+1994,Rothstein+2008,Zhu+2018}.
If disks are turbulent, then the  large scale magnetic field  
is subject to  turbulent diffusivity  just like the velocity.  The implications of standard
accretion theory being a mean field theory are underappreciated in this context \cite{Blackman+2015}.
A complete accretion theory should involve coupling the mean field
velocity and mean magnetic field, the latter of which then incurs both  advection, diffusion, and dynamo growth coefficients.  Accretion of flux through a thin disk is suppressed with respect to advection because the
field diffuses from a term with gradients associated with the small vertical variation scale whereas advection occurs as a diffusive velocity associated with variation on scale $R$.   

Advection might be effective over the surface layers of a disk.  \cite{Rothstein+2008,Zhu+2018}.
but there must be an accumulation of an average net sign over a period long enough to power 
observed jet durations.  The question  remains as to what  supplies and determines the coherence
time of the signed flux. Ultimately, the answer has to be a dynamo, but whether the dynamo 
is in the disk  supplies flux
that is accreted from the ambient ISM remains an open question.

A common pitfall of both jet and accretion simulations  is that answers to posed questions can
inextricably  dependent upon the tool and initial setup. Traditional  jet simulations from which we have learned about the basic launch and collimation,  impose the initial magnetic field and treat the disk as a boundary condition, e.g. 
\cite{Pelletier+1992}.
Even when an accretion disk is treated with global MHD, reconciling the  connection between
 the  initial field geometry and the end state is an ongoing enterprise
\cite{Penna+2013,Hawley+2015,Zhu+2018,Liska+2020}.    In some GRMHD simulations that start out with toroidal field 
or mesoscale poloidal loops in a finite mass disk, the end state seems to converge to a state with a large scale poloidal field. field \cite{Penna+2013,Liska+2020}, suggesting there maybe  some independence of   initial state.
This suggest a  dynamo and magnetic relaxation toward the relaxed state of  helical field is at work. However,  a real disk may have a steady source of  mass resupplied  and an actual steady-state would represent some not-fully-relaxed state that represents  a competition between relaxation and turbulence.

Cartesian  box simulations  of "accretion" commonly invoke shear periodic 
boundaries in radius and azimuth, and sometimes vertical as well e.g. \cite{Balbus+1998}.
 They have no mass transfer in the disk and thus no actual accretion and  no actual angular momentum transfer either, just linear momentum, since the radius of curvature is infinite.
 Quantities averaged between periodic surfaces cannot change,
in contrast to a  local Cartesian section of  a real disk. 
Most importantly, 
real disks also involve large scales not contained in such local boxes, and if large scales including jets e.g \cite{Konigl1989}
 or corona \cite{Lynden-Bell1969} dominant transport (see \cite{Blackman+2015} for further discussion), 
  boxes   won't capture the dominant processes. 
Global disk simulations are now increasingly common, but there are  challenges
for convergence  \cite{Hawley2011}, not only in stress magnitude but also in the stress  spectra that
reveal which
  transport scales dominate.
%
The extent to  which convergence depends on  initial conditions like the initial geometry of the magnetic field and role of finite initial mass also remains to be understood.

We come  back to the plausible paradigm that accretion disks  sustain both small and large scale  fields and those of sufficiently   large  scale   buoyantly rise to the corona without shredding from turbulent diffusion.    Such buoyant loops should then be recognized as loops of flux or ribbons, rather than lines, to properly account for magnetic helicity.  Once in the magnetically dominated corona, continued foot-point motions twist the field ribbons further  and inject magnetic
helicity, possibly with different signs on different scales  \cite{Blackman+2003}, not unlike what is observed sun \cite{Pipin+2014,Singh+2018}. In response, the loops can kink or buckle and reconnect.  In doing so, some  structures dissipate while some relax to larger scales, minimizing their energy for a given magnetic helicity. The energy lost in the relaxation becomes coronal emission and  the opened fields may be the source of the large scale fields needed for jets. 


\subsection{How do large scale fields actually mediate jet launch?}

Though virtually all jet models that appeal to large scale fields  involve some  axial magnetic pressure gradient and collimating toroidal magnetic field, 
 there are basically three classes of magnetized models that observations have not 
universally  been able to distinguish.  Which, if  any of these are these plausible
models operate in a given system is  an open question.   The discussion below
is dominated by insight and interpretation   from   theory and simulation. 

\subsubsection{Magneto-centrifugal (MCL)}

The standard MCL model is usually presented  as a steady-state \cite{Blandford+1982}, model with 
large scale poloidal magnetic field having  one set of foot-points anchored in the disk and the other  at infinity.  The field is quasi-rigid from the disk surface out to the Alf\''en surface associated with the jet flow.  Within the Alf\'en radius,  the field moves quasi-rigidly, being anchored to the rotator.
As such, points on the quasi-rigid field  farther  from the source than their anchor points have  super-Keplerian rotation for their location. For those field lines (or rather, field bundles which allow for magnetic helicity) inclined less than 60 degrees to the outward radial vector, a net force along them results,   accelerating the mass loaded thereon.  In a steady-state,  mass continuity may require mass to be initially  supplied thermally, above which  a centrifugal "fling"  takes over, Near the Alf\'en radius,  the field  ratio of  toroidal to poloidal field  approaches unity. There the toroidal magnetic pressure gradient takes over  the acceleration  and   hoop stress collimates. Eventually by $\sim 100$  gravitational radii or so, the flow accelerates to the point that its ram pressure dominates.  In a rotating frame, the dominant force has therefore  transitioned from  centrifugal,  to magnetic,  to ram pressure by this stage.   For AGN this would happen well below parsec scales. 

The MCL can also emerge from a non-steady state initial condition with say a unidirectional
 vertical large scale magnetic field that decreases in radius. \cite{Pelletier+1992}.  Then the inner field can bend the outer
 field lines into the critical angle and an MCL will arise self-consistently.  
 Nevertheless, the model always requires the existence of  a large scale ordered field.

\subsubsection{Blandford-Znajek (BZ) mechanism}

This  mechanism \cite{Blandford+1977,Penna+2013}
 appeals directly to  rotational energy extraction from a black hole ergosphere threaded by a magnetic field 
whose open field lines are  anchored  at infinity.  This model is  like  a  black hole analogue of a pulsar. \cite{Goldreich+1969}
The mechanism works only for  a rapidly spinning black hole supplied by magnetic field from the disk, and a pair plasma
cascade from disk gamma ray annihilation to to carry currents. 
The source of the magnetic field here again an open question. 
To avoid field annihilation by reconnection, there must  be sufficient net flux of one sign over the lifetime of a jet, $10^8{\rm yr}$  in the case of AGN, which is orders of magnitude longer than 
accretion time scales of the inner disk.   
The BZ depends on the disk both for its magnetic field and likely for supplying   angular momentum  to the black hole.  Poynting flux dominates the jet launch at the base.

This mechanism could coexist with an MCL outside of it  in black hole engines, and simulations have studied the combination of the two. \cite{Ferreira+2013}.  Perhaps the MCL could even help collimate/stabilize the inner  BZ jets.  The MCL depends only on the disk and could operate regardless  of the black hole spin, although a rapidly spinning hole means the disk extends deeper in the potential well and even the MCL outflow  becomes more powerful.   One key difference between  the MCL and the BZ is that matter content in the former is disk plasma, composed  of ions and electrons, whilst the BZ plasma is pair plasma. If the composition at the base can be determined, this could be  help distinguish the two.  We come back to the jet composition  later. 

\subsubsection{Magnetic Tower (MT)}

The magnetic tower \cite{Lynden-Bell2003,Uzdensky+2006,Li+2006,Huarte-Espinosa+2012,Gan+2017}
is a third paradigm for magnetically mediated outflows which  is also Poynting flux dominated near the base.  This involves no centrifugal launch, but  rather poloidal magnetic loops with   foot-points anchored  in surfaces that are in relative differential  rotation. Most naturally this could be foot-points  linking   the central object with the disk.  Differential rotation causes the field to wind up and grow a toroidal component.  The resulting gradient in toroidal magnetic pressure pushes up the magnetic structure. The rising magnetic tower is mostly force free, except at the very top where material above the tower is pushed ahead.  Importantly, unlike  BZ which has only one sign of flux in in the jet,   the MT  has both the upward flux and the downward flux within the jet tower, so that jet has no net vertical magnetic flux in either hemisphere.  Also, when treated  overly simply as a series of equilibrium force free states 
 in the absence of an ambient pressure, the   mechanism  produces only a broad splayed wind \cite{Lynden-Bell+1994}. 
 
MT collimation in the treatment of  Refs. \cite{Lynden-Bell2003,Uzdensky+2006}  
depends on the ambient pressure.  However, enhanced self-collimation even without ambient pressure might be possible  \cite{Bellan2018} as discussed further in section \ref{coaxial},  although  stability  over large distances may still require an ambient plasma.  In this respect,  magnetic jet engines within stellar envelopes are quite favorable circumstances, as may arise for long gamma-ray bursts
 \cite{Uzdensky+2007}, or post AGB stars/planetary nebulae  \cite{Vlemmings+2006}.
The MT also has a highly collimated spine of flow or particle dominated  plasma where the toroidal field becomes small, while surrounding this core is the magnetically dominated structure.  The plasma $\beta$ of the MT jet  therefore   exceeds 1  at the core, and drops below 1 away from the core.    

Compared to the MCL which becomes asymptotically flow dominated, the MT can remain super-magnetosonic out to larger distances, possibly all the way to the radio lobes in the case of AGN \cite{Gan+2017}.  Like the  MCL, but unlike  BZ, MT field lines (or field ribbons)  have one foot-point on the disk, which has ion-electron plasma. 
The plasma of the MT could therefore have a higher fraction of    ion-electron plasma at the base than  BZ, but perhaps less than MCL,  highlighting again the potential importance of the compositional differences in distinguishing  different paradigms.

Unlike the MCL and BZ paradigms, the MT does not need  a global scale field as a starting point.
Rather, a  mesoscale field that links the disk to  central body will do.  Since this circumstance is likely rather generic, the MT potentially offers a simpler starting point, although a coherent sign of the flux may be needed to avoid premature dissipation of the tower by  reconnection between azimuthal loops.
Note also that the MT is typically presented in a time-dependent picture, whereas  standard approaches to the MCL and BZ  focus on  a  steady-state picture, requiring  mass  loading onto field lines.
This subtle difference can cause confusion by hiding the fact that all three mechanisms do ultimately depend on a  vertical component of gradient in toroidal magnetic pressure to push material  ahead.   All three mechanisms also have  regimes of magnetic domination and flow domination. For the MT, the latter is at  the core and at the very top of the tower but  most of the MT is magnetically dominated. For the MCL, most of the jet becomes flow dominated on on observable scales outside the very core. 

Despite subtle predicted theoretical differences, it remains difficult to   confirm observationally the relative importance of BZ vs MCL vs. MT.
So far, even the  (EHT)  \cite{Jimenez+2018} does not really distinguish these models, although it at least shows evidence for some kind of ordered magnetic fields mediating the outflows at the base. 

The fact that the MT requires only initial mesoscale fields and does not depend on centrifugal forces, also makes it more amenable to testing in  laboratory experiments as discussed later.

\subsubsection{How do explosive jets work?}

Since the engine time scales are so short for jetted stellar mass engines,  
 jets of GRB and  SN  \cite{Wheeler+2002} that may be 
 quasi-steady  on orbital time scales of the central engine, 
  still appear  explosive to the observer.  Both MCL and MT type models could then apply.

But another  transient magnetic outflow model is a  magnetic bomb (MB) \cite{Matt+2006} 
\cite{Wheeler+2002}.
Like the MT and the MCL, the  MB thrives from  gradient of toroidal magnetic field pressure, but it  is highly time dependent in that a  jet may result only after sufficient magnetic field is wound  up by a newly formed rotator below.  Instead of the time evolution of the system being a steady progression  though a series of near equilibrium states,
the field may need to be wound up sufficiently, and only then "explodes".
Ref.   \cite{Matt+2006}  showed, for both quadrupole or dipole fields on a rotator weighed down by an ambient medium,  that once the field winds up to a critical threshold, the  bomb accelerates material both axially   and equatorially. 
The vertically  decreasing gradient in toroidal magnetic pressure causes the  polar outflows, but since   the toroidal magnetic pressure peaks away from the mid-plane in each hemisphere,   material  is also pushed  toward the mid-plane by both hemispheres and   squeezed out along the equator.   Some variation   could play a role  in SN 
 in gamma-ray bursts (GRB), but the concept was used originally to help explain features of planetary nebulae morphology, including  ansae \cite{Stanghellini+2016}.

\subsection{What  turns jets on and off?}

Of the jetted sources, x-ray binary microquasars  \cite{Fender+2004,Ferreira+2006,McClintock+2006,Corbel2011,Kylafis+2015}
are particularly intriguing because not only do they involve compact, stellar mass scale compact  relativistic  engines within our galaxy, but the orbital time scales at the inner disk radii are of order milliseconds.   Typical X-ray observations of these sources  last    $>10^6$ orbital periods,
 and therefore provide a very comprehensive ensemble averaged time evolution of these systems. 
Since they incur  outburst states lasting from $\sim 20$ days to many months   \cite{Remillard+2006} depending on the object,  and  transition  to quiescent states, we  have a  comprehensive record of their transitions over decades of observations.  The quiescent  or "low" states have harder spectra than the more luminous  "high"   states and so the  monikers "high-soft" and "low-hard"  give a broad  characterization.   Jets are prevalent in the low hard state, which are radio loud.  Do these states also tell us about circumstances in which jets form and the corresponding evolution of AGN as well? Since the orbital time scales, and likely the  state transition time scales for AGN are 7 or 8 orders of magnitude larger we cannot generally detect their state changes in real time.

The state transitions likely result from changes in the accretion rate, and one 20th century paradigm to explain the transition from the high-soft to low-hard state  is a transition from thin disk to thick advection dominated accretion flow (ADAF)  \cite{Esin+1997},  a  sub-class of Radiatively  Inefficient  Accretion Flows (RIAFs) \cite{Yuan+2014} with conceptual origins dating back to early two-temperature disk models \cite{Shapiro+1976}.  In this paradigm, when the accretion rate drops sufficiently, the collision time between ions and electrons becomes so low that  the accretion time scale is shorter than the time for electron-ion equilibration.   

 Two  plasma physics assumptions  then become essential:  (1) the transport of angular momentum occurs via a turbulent viscosity that dumps the dissipated energy directly into ions and NOT electrons, and (2) there is no faster-than-Coloumb coupling betweens ions and electrons.   Whether the aforementioned plasma physics assumptions hold may be amenable to  laboratory experiment, although none has yet been directly performed. 
Whether the above assumptions are valid have been topics of active research \cite{Begelman+1988,Quataert+1998,Blackman1999,Pariev+2005,Sharma+2007,Park+2010b,Sironi2015,Zhdankin+2019}.  If they do  apply, 
then  the weakly accreting disk puffs up as the ions carry the dissipated energy and  advect it as internal energy. 
 Since they accrete to the central object before transferring their energy to electrons, the radiative efficiency is weak,
 explaining the quiescent states.  
  Interestingly,   the jets are more prevalent  in the quiescent states  \cite{Fender+2004,Kylafis+2015}, suggesting that a large thickness of the disk may play a role in collimation. 
  
Since AGN do not accrete in binaries, but from an ambient medium, the temporal pattern of  accretion rate variability may not be directly analogous to that of microquasars. But the consequences of varying accretion rates may be analogous. Most galaxies, despite having massive black holes at their cores, are in rather quiescent accretion states. ADAFs in their original form may not be sufficient, particularly if electrons are directly accelerated by angular momentum transport dynamics, and so  mass loss via winds
\cite{Quataert+1999}, or convection dominated accretion flow RIAFs 
\cite{Narayan+2000,Quataert+2000} may be important.
 Nevertheless the microquasars do suggest a direct connection between thick disks, hot ion tori, and jet collimation and this connection  arose early in theoretical work on AGN  \cite{Rees+1982}.

\subsection{Are relativistic jets  launched by a fundamentally  different mechanism than non-relativistic jets?}

Having distinguished   launch paradigms, we can ponder
whether the dominant  launch mechanism differs for 
relativistic  vs. non-relativistic jets.   Do the relativistic jets of AGN and microquasars simply result from  deeper potential well versions of whatever produces jets in YSO \cite{Price+2003}? 
If so, then this would imply that jet depends primarily  on the disk and relative motion between the disk and central object, and maybe less so on the nature of the central object and the general relativistic physics, or pair plasma associated with strongly gravitating engines.
Debate on this issue began as soon as jets were confirmed in both relativistic and non-relativistic sources
and persists, given the ubiquity of jets from all types of objects.

\section{Propagation Region Mysteries}

Even if jets have Poynting flux dominated regions within $\sim 100$ times the inner disk radius of their base,  most observations probe their  much  larger propagation scales. 
The questions of whether jets require active collimation during  propagation,  
whether  magnetic fields are dominant or subdominant, and what  the matter and energy are composed of
 persist.   We discuss each below. 
 
\subsection{What   jet collimation mechanisms operate in nature?}
\label{collimation}

Although magnetically mediated jets are often thought to be "magnetically collimated," there is a significant difference between launch and collimation  in practice.  There can also be confusion  by what is actually meant by "collimation."
For example, is an ambient medium required to  stabilize an otherwise magnetically collimated structure?  In that sense, the magnetic field  may collimate  a flow-dominated regime along its spine,  but the ambient medium may prevent kink or firehose instability from completely destroying  the jet coherence. In that sense, both the magnetic field and the ambient medium are needed to explain the observed structure. 

From the jet aspect ratio alone, it is difficult to determine which of the following dominates  collimation:  (i)  accumulated  magnetic stresses over the course of  the propagation; 
(ii) collimation   simply from a collimated super-magnetosonic  launch such that the aspect
ratio is  determined by the inverse tangent of the ratio of 
expansion speed to jet flow speed. This may be further aided by cooling  \cite{Stone+1993,Lebedev+2002}; 
(iii) inertial focusing by a surrounding wind or ambient medium.  In fact, there exists no jetted source where the absence of a surrounding wind or plasmas is proven absent.
There is  direct evidence  for jets surrounded by  winds  in AGN \cite{Yang+2012}, YSOs \cite{Arce+2002}, pulsars \cite{Kargaltsev+2003}, or embedded in stellar environments, including post AGB and planetary nebula \cite{Imai+2002,Vlemmings+2006,Witt+2009}, and SN and GRB \cite{Wang+2008,Aguilera-Dena+2018}.

In assessing  magnetic collimation mechanisms, we reiterate that 
MCL jets are  flow dominated outside of  $\sim 100r_g$,  but  the lateral hoop stress of the field can still be influential over large distances. MT models can in principle  be magnetically dominated   out to the hotspot  in AGN jets \cite{Colgate+2014}.   Simulations have shown that a MT can remain  at least modestly magnetically dominated  out to $1000r_g$ \cite{Gan+2017}.
Observational evidence suggesting magnetic domination of AGN jets out to very large scales
 is the fact that some jets   show nearly right angle bends  on kpc scales \cite{Eilek+1984,O'Donoghue+1993}.
 Purely hydrodynamic  jets cannot easily incur such extreme bending  and remain collimated. MT  fare better \cite{Gan+2017}.   A caveat may be that  what appears  as a bend could  somehow just be a  pattern of radiating particles.   The jet  viewing angle may also deviate  significantly form a face on view  such that a gentle bend looks sharper when projected onto the sky, or interacts  with an ambient medium \cite{Rawes+2018}.

Evidence for ordered helical fields in jets on large scales via  polarization and Faraday rotation has also been interpreted as evidence for magnetically dominated jets on large scales.
\citep{Lyutikov+2005,Asada+2008,Gabuzda+2018}.  A caveat  is that turbulence does not diffuse helical mean  fields  efficiently, even if their energy density is subdominant  \citep{Blackman+2013,Bhat+2014}.  This is because conservation of magnetic helicity energetically disfavors diffusing that magnetic helicity to small scales.   Ordered helical mean fields  could persist  even if the small scale field is tangled and  so the spatial resolution of observations becomes important in assessing what is actually being measured.

\subsection{What is the jet particle and free energy composition?}

For jets from central engines other than neutron stars and black holes,   the only plasma available is ion and/or molecule rich.   But for black hole engines which  power AGN and microquasars, 
 the question of composition is  germane because  different launch mechanisms 
 involve  different  compositions.  BZ predicts jets launched with 
 primarily pair plasma,  whereas the magnetic tower and BP include more
ion-electron plasma.  For black hole engines, mechanisms to load the field lines with plasma and the content distribution remain ongoing areas of research  \cite{Romero+2020}, but there are some earlier notable efforts.   Combining radio spectral data,  predictions from  synchrotron-self absorption, 
and kinematics  \cite{Reynolds+1996} argued that M87 is likely pair plasma in the core on scales below 0.01pc.
 The same conclusion via a different analysis was reached in Ref. \cite{Pashchenko+2019},
for  NGC 315.  Care has to be taken in assessing key assumptions of the underlying jet models, specifically hidden assumptions about   the ratio of magnetic  to particle energies. 

The  absence of Faraday rotation for a pure pair plasma is another  possible method to constrain composition.\cite{Park+2010a}.  Equality of positron and electron masses imply that left and right handed polarized waves propagate at the same speed.  If independent constraints on magnetic field strength and lepton density are available, then the lower the Faraday rotation, the higher the pair plasma  content.  
There are complications in practice. For relativistic flows, there is degeneracy  in  field geometry,
jet orientation, and Lorentz factor for  given values of  rotation measures at a given field strength \citep{Lyutikov+2005}.  In addition, ambient near the  jet boundary could dominate the rotation measure . Also, farther from the  engine, even more mixing between jet and ambient material, or baryon mass loading from wind emitting stars moving into the jet e.g. \cite{Komissarov+1994,Hubbard+2006} 
make distinguishing entrained jet material from launched jet material rather challenging.

Other indirect methods to determine composition include using statistics of neutrino detections 
\cite{Garrappa+2019}.
If they  are  associated  with pion production via  proton-proton collisions, then their detection constrains the  proton fraction in blazars when combined with other measures of  inertia.
From spectral modeling, \citep{Celotti+2008,Ghisellini2014} argued that blazar data favor protons as carrying the bulk of the momentum, within the assumptions of their emission model which presumes a  tangled magnetic field, steady-state, and synchro-Compton emission.  Taken at face value, these  results  contradict conclusions reached above that M87 and NGC 315  are pair plasma dominated if in fact classification of  radio jets is essentially determined by orientation \cite{Urry+1995} with blazars being the class of powerful jets   viewed end on.  This contradiction gives a flavor of the challenges
and  open questions regarding  composition  determination and highlights  the  need for a systematic review and new methods. Might these include laboratory experiments?

\section{Jet Dissipation and Particle Acceleration  Mysteries}

Taken collectively, jets are not uniform in emission. In some jets,  for example FR2 sources,
emission is largely absent  until the hotspot or radio lobe. In others, such as FR1 sources, 
the jets are bright along the length of the jet.  Some like M87 exhibit  knots or bright spots within their collimated structure.  Ref. \citep{Matthews+2020} discusses  particle acceleration with a specific focus on jets in this context.
 
\subsection{What causes jet Instability and local sites of particle acceleration?}

To produce local sites of emission within jets,  there must be a mechanism to convert some bulk free energy   into particles. If the jet is flow-dominated, then shocks  are  most natural.   They have been widely invoked to explain knots and hotspots. \cite{Bicknell+1996}.   If however, the flow is magnetically dominated,  then magnetic reconnection sites are plausible.   To produce either shocks or magnetic reconnection  some kind of obstacle or instability is required.  For shocks, the jet launch could be unsteady,  turning on and off.  In the specific case of AGN, a wind-emitting giant star passing into the jet could also produce flow that triggers shocks \cite{Hubbard+2006}.    For magnetically dominated jets, the kink instability is a natural mechanism   to trigger  reconnection, when the system tries to evolve back toward its relaxed state. \cite{Meier+2001,Tchekhovskoy+2016}.

Since the kink instability  occurs for sufficiently twisted magnetic flux bundles,  determination of whether  helical structures in jets  are magnetic or hydrodynamic  has long been of interest. 
A potential observational prediction that emerges from simulations is that propagation of hydrodynamically produced  helical  structures  would involve wave propagation of that structure  along the jet \cite{Hardee+1992}
whereas  the magnetic case would involve the physical flow following a helical path \cite{Nakamura+2004}.


Particle acceleration at collisionless shocks as a means of dissipating hydrodynamic flows has  been studied for decades in relativistic and non-relativistic flows, with a variety of analytical and numerical
methods (for reviews see Ref. \cite{Treumann+2009,Matthews+2020}).   Different methods  have  strengths and weaknesses. For example, self-consistent generation of shocks from the microphysics  occurs in particle in cell (PIC) simulations, which  can also track the detailed mechanisms and conditions for ions and electrons to be accelerated.  However PIC simulations are also limited to very small  scales compared to astrophysical jet scales, and maybe even  small compared to the entirety of Fermi-type diffusive shock acceleration scales.   Test particle, and other more macroscopic or semi-analytic approaches, have the complementary problem of not capturing the true microphysics from first principles.   Ultimately, one hopes that the output microphysics simulations can be used to validate more practical semi-analytic approaches for direct astrophysical modeling.   

Complementing  shock acceleration  is  magnetic reconnection.   This has  long been a separate topic of active research with its own active community.   Understanding/determining the rate of reconnection 
 and how  converted energy is partitioned into  flow and particle energy spectra have been   the 
 traditional challenges.   Through a combination of theoretical, numerical and experiments, the rate of reconnection is now    much better understood  \cite{Zweibel+2009}.For a wide range of astrophysical plasma conditions the speed of reconnection  seems to be consistent with  $\sim 1/10$ the Afv\'en speed in magnetically dominated plasmas.   Turbulence is likely to play a dominant, if not simplifying, role in facilitating fast reconnection \cite{Lazarian+1999, Blackman+2008,Eyink+2013,Lazarian+2012,Lazarian+2020}, in the larger scales of real astrophysical  reconnection sites compared to what is possible to simulate and measure
 experimentally. 
 
  Predicting and understanding  accelerated  particle energy spectra from reconnection  is the primary frontier. As a particle accelerator,   magnetic reconnection should really be thought of as an acceleration environment, not a single mechanism.   The extent to which electrons vs. ions get accelerated  and which mechanisms operate in a reconnection region
  can depend on plasma conditions, boundary conditions, and system size \cite{Blackman+1997,Workman+2011,Fox+2017,Che+2019}.
Mechanisms operating in reconnection environments may  include some combination of direct electric field acceleration, stochastic Fermi acceleration from turbulence,  first order Fermi  acceleration from converging flows or within  magnetic blobs, and downstream fast shocks.   One way or another, turbulence is also likely an important  player for particle acceleration in reconnection regions e.g.
 \cite{Larosa+1996,Selkowitz+2004,delValle+2016,Phan+2018,Guo+2020,Ergun+2020}. As with shocks, scale limitations of simulations here too must be kept in mind. While efforts have sometimes focused on specific structural features that develop in reconnection simulation outflows, e.g.  Ref.  \cite{Sironi+2016}  for relativistic pair plasmas,  it could be that Lundquist and magnetic Reynolds numbers for realistic  systems   are so high that the full reconnection region becomes fully turbulent.
If the end state  becomes  largely independent of the preceding details of the initial instability  or intermediate states, this would be a  practical conceptual simplifciation.
 
 One overall message from the state of present simulations,  is that 
relativistic electron acceleration acceleration, at least in relativistic pair plasma simulations seems to    efficient   \cite{Sironi+2014,Werner+2017,Guo+2020}, but this  all remains an active enterprise of  research,
including the role of simulation box conditions and the role of ions.

\subsection{What causes continuous emission within jets?}

A difficulty of  particle acceleration  in AGN jets and lobes is that the synchrotron loss time is often much shorter than the flow crossing time of the emitting region \cite{Meisenheimer+1986}.  Thus, even if local sites of acceleration are efficient, there needs to be widespread re-acceleration.   This becomes more natural if there is turbulence so that wave-particle scattering can sustained, for example by stochastic Fermi-type acceleration.   Turbulence in a flow dominated jet might arise from strong shear flows \cite{Liang+2013}. 
If the jet is magnetically dominated  such as  in a MT, jet  reconnection sites that accelerate particles \cite{Blackman+1996}  could arise from  kink instabilities, which  can  also facilitate the development of 
or a  spectrum of plasma waves   that   accelerate particles  via resonant or non-resonant scattering
without destroying the overall collimation of the jet  \cite{Colgate+2015,Tchekhovskoy+2016,Alves+2018}.
Jets driven by tangled magnetic fields \cite{Heinz+2000} might also be possible, in which case the tangled field
in dynamical steady steady-state could incur  dissipation throughout the jet, and the magnetic field would be  resupplied from below.  

As to what distinguishes FR1 and FR2 sources in this context,  the debate has long been whether the
distinction  results from different engine powers  or different environments \cite{Gopal-Krishna+2000},  including
stellar wind mass loading \cite{Perucho+2014}.
The two paradigms may not be entirely decoupled. There may be a continuum of engine powers 
and also   a threshold gas pressure of the ambient environment for a given power that determines the distinction. Ref. \citep{Tchekhovskoy+2016} showed  that MT-type jets with low enough power 
running into an ambient medium with high enough pressure 
can decelerate the jets  enough such that toroidal field  piles up above the critical threshold for kink instability.  Such kink unstable jets become the FR1 type,  whilst FR2 would be those jets with higher power that  penetrate through the ambient medium without becoming kink unstable.  This is also consistent with the fact that  FR1 appear mostly in   cluster ellipticals with a high ambient gas content, whereas FR2 come from non-cluster ellipticals. 

Shear between the jet flow and  the ambient medium can also generate turbulence at the jet boundary.
  Acceleration and  emission would  not  be edge filling, but appear space filling only in projection on the sky.
 The jet could  be hollow in particle acceleration.   It is hard to assess observationally, even  for blazars, whether jet emission is primarily an edge effect.  
 Particle acceleration from  relativistic shear  flows has been studied to some extent at the PIC level
\citep{Liang+2013} and a Ref  \citep{Rieger2019} provides more recent pertinent 
review of shear flow acceleration. 

The extent to which jets are fully turbulent remains a key question for diagnosing particle acceleration and theunderlying mechanism of launch.  Importantly, the flow can have both an overall mean set of properties,  and fluctuations and turbulence superimposed.  We again emphasize that the presence of ordered mean properties does not preclude the contemporaneous presence of influential  turbulent fluctuations.
 
\section{What have we learned from experiments so far?}
 
Astrophysical objects cannot be produced in the laboratory.  Even when physics regimes
can be scaled to laboratory conditions in certain key parameters \cite{Ryutov+1999,Ryutov+2000},  the boundary conditions and the dynamic range are still off by orders of magnitude.   Were the aim 
to produce complete objects or jets from launch to dissipation over realistic dynamic ranges, 
jet laboratory  astrophysics could not be well justified. Realistic  useful goals  do however   include one or more of the following for these demanding experiments:  proof of principle of a physical process  under  somewhat controlled conditions that are otherwise relegated only to numerical simulations and theory; 
benchmarks for numerical simulation; measurement of   thermodynamic relations and transport 
coefficients that only elsewhere occur in astrophysical objects;  discovery of  new, if not  unexpected, physical phenomena;   thoughtful approaches to  progress toward solving  an  astrophysical mystery. 

In the previous sections we toured some  long standing   jet-related  physics  questions that have persisted for decades
to set the context within which to assess  and challenge laboratory jet experiments.
In this section  we summarize some of  what has been accomplished in  laboratory astrophysics experiments that connects to jets from engine to dissipation, and the physical topics indicated by the subheadings.
We keep in mind the separate physics questions within  each of the aforementioned  jet regimes:
engine/launch; propagation/collimation ; dissipation/particle acceleration. 
So far,  experiments  have  been mostly limited to the "proof of principle" level, albeit with some notable elucidation of jet collimation physics.    There remains a significant gap (and thus opportunity) 
between lessons learned from these experiments and direct solution of the  challenging open questions discussed in the previous sections. 

\subsection{Dynamo and MRI}

As described above, one way or another dynamos are important to the origin of magnetic fields in the engine.
In discussing relevant experiments, note that dynamo means different things to different communities and it is important
to clarify definitions when discussing what different experiments measure.
At its most basic level, we can define dynamo simply as  {\it exponential amplification of magnetic energy  within some range of scales that sustains against exponential decay}.  Inside astrophysical rotators,  dynamos  feed on some combination of kinetic energy in  turbulence and  rotation. Large scale dynamos are those in which the scale of the kinetic energy in field growth occurs on scales large compared  to  turbulent motions. Small scale, or fluctuating dynamos are those in which the magnetic energy occurs at or below the scale of dominant turbulent motions.  Theory and many simulations have found that the MRI is a source of both large scale and small scale dynamos operating contemporaneously \cite{Brandenburg+1995,Lesur2010,Davis2010,Simon2011,Guan2011,Sorathia2012,Suzuki2014,Shi+2016,Bhat+2016}. 
 The MRI  actually operates first as a large scale dynamo,  before mode coupling produces small scale amplification, and then the saturated state has some combination of the two \cite{Bhat+2016}.  

Large scale dynamo experiments, small scale dynamo experiments, and MRI experiments have all been pursued
separately.  Large scale dynamo experiments using liquid metals have proven that 
large scale field growth is possible when the system is subjected to  helical forcing e.g. \cite{Gailitis+2004,Gailitis+2018}.
These impressive proof-of principle experiments are however, at rather low magnetic Reynolds rather than  the turbulent  regime of astrophysical rotators. Also, their flows are  imposed with the required pseudoscalar
properties, unlike  astrophysical circumstances where the flows purportedly  emerge naturally from e.g. density stratification and rotation.
 The experiments also  do not address  how large scale dynamos saturate in turbulent flows, but they
can address saturation by  differential  rotation saturation e.g.  \cite{Gailitis+2004,Gailitis+2018,Stefani+2018}.
Many  numerical simulations also focus on highly idealized version of dynamos
and so the experiments are not alone in simplifying the astrophysical circumstances. There is some art to understanding which simplifications are acceptable. In any case comparison between simulations and experiment is itself valuable in these contexts.

 Small scale dynamo experiments now exist using laser driven plasmas \cite{Tzeferacos+2018,Bott+2020}.  The concept here is that laser driven ablation of metal foils are arranged so that plasma is guided into plumes that collide to produce
 a turbulent region of high magnetic Reynolds number. Amplification of the magnetic field is measured via 
 faraday rotation and/or proton radiography. The latter techniques are subtle and so far require assumptions about the statistical properties of the flow and magnetic field turbulence.  These experiments do seem to show total magnetic field amplification consistent with an MHD small scale dynamo.   Measuring the spectra is still in its  first generation of diagnostics,
 and has a limited dynamic range, so is likely   subject to further refinement. But this does represent the  emergence of a potential tool for more detailed symbiosis  of theory, simulation and experiment.  
 These small scale dynamo experiments are aimed at comparison to theories of isotropically  forced systems
 e.g \cite{Schekochihin+2002}, although  they use interacting plumes to  approximate  this state. 
 The extent to which different forcing methods and geometries  lead to similar or different results and better or worse homogeneity in the turbulence would be   an interesting direction  for further work. 
 If the experiments reach the point where the magnetic spectra can  distinguish helical vs. non-helical forcing, that would also of interest to compare with theory and predictions of large scale field growth.
 
Except for the  potential role of star-disk collisions in AGN \cite{Pariev+2007a,Pariev+2007b}, dynamos in accretion disks are likely forced by differentially rotation.  In a steady-state the MRI and its associated dynamos are fully non-linear and saturated.  There have been efforts to measure the  onset of the "standard MRI" (referring to case of initial vertical field) in the laboratory 
using liquid metals  \cite{Sisan+2004}, but   complications associated with purely fluid  effects such as Ekman circulation challenge this  interpretation \cite{Schartman2012}.
The mechanical analogue of the  standard MRI mechanism has  however been demonstrated experimentally
\citep{Hung+2019}.  The inductionless low magnetic Reynolds number helical MRI or toroidal field MRI 
has  been measured in liquid metals \citep{Stefani+2009}.   All of  these  experiments remain at the level
of proof of principle. They are not yet tools to explore the detailed saturation properties, nor the actual transfer of angular momentum since there is no central gravity source and no mass flow
as in a real accretion disk.   They  also cannot  connect the MRI to advection or generation of the fields needed in the engines that buoyantly rise to coronae and relax to become jet-mediating large scale fields.  

 The  definition of a  dynamo above also accommodates magnetic relaxation--whereby  electromagnetic energy and magnetic helicity is injected into a system on small scales, but relaxes to large scales.  Relaxation tries to bring the system to a state of minimum energy but the electromagnetic  forcing on small scales competes against the fully relaxed state. At the same time, this forcing stresses the system to be unstable which in turn sources fluctuations.  An EMF arises from the mean cross product of  velocity and magnetic  fluctuations.    This is the  same kind of term that drives  large scale magnetic field growth when fluctuations are sourced by forcing of kinetic energy in flow driven astrophysical dynamos.   This process of magnetic relaxation can be thought of as a magnetically driven dynamo
 and has long been studied in the context of Reversed Field Pinches and Sperhomaks \cite{Strauss+1985,Boozer+1986,Bhattacharjee+1986,Ortolani+1993,Bellan2000,Ji+2002},  and plays a role in the Spheromak jet experiments described below. A quasi-steady state can arise with steady injection, possibly with sawtooth oscillations.  
  
The  magnetically driven dynamo does have a direct and important analogue in astrophysics, namely   the injection of   buoyant magnetic structures into coronae. Note that for all objects but the Galaxy, observations only measure the magnetic fields  exterior to the bodies, usually in magnetically dominated corona. Any  complete  astrophysical  dynamo theory must really connect the interior flow dominated dynamo  to an exterior magnetic relaxation dynamo.   This connection is  likely important for allowing the needed flux of magnetic helicity out of the interior in to sustain rapid cycle periods of stellar dynamos in the view of dynamos that emphasizes the essentiality   of magnetic helicity fluxes \cite{Blackman+2000a,Vishniac+2001,Blackman+2003,Shukurov+2006,Blackman+2015}.

A detailed discussion of  open questions and controversies in dynamo theory are beyond the present scope,  but  some recent reviews on various aspects of the subject  include Refs. \cite{Brandenburg2005,Blackman+2015,Rincon+2019,Tobias+2019}.

\subsection{Magnetic Reconnection and Shocks}

Magnetic reconnection experiments are reviewed  elsewhere \citep{Yamada+2010,Yamada+2016}.
They have been helpful  in  understanding aspects of the rate of reconnection and the regimes of slow
vs. fast reconnection.   Low plasma $\beta$ reconnection regions of 
sufficiently large Lundquist number $\ge 10^4Pm^{1/2}$  \cite{Baalrud+2011,Loureiro+2013}
where $Pm$ is the magnetic Prandtl number, 
develop  instabilities such as e.g. the plasmoid blob  instability  that allows flux dissipation with an inflow
speed to the   reconnection  layer  $\sim 0.1 v_A$ where $v_A$is the Afv\'en speed associated with the pre-reconnection flow.  For experiments and simulations below  the critical   Lundquist number, the rate of reconnection follows more closely the Sweet-Parker rate, which  varies with the magnetic Reynolds number as $R_M^{-1/2}$.
It is likely that in real astrophysical environment with much larger Lundquist and magnetic Reynolds numbers however, that  the plasmoid blob instability is just the tip of the iceberg and that the flow past the blobs itself may drive shear
instabilities that causes fully developed turbulence. 
Then the details of the  instability are  much less important than what may be a ubiquitously turbulent end state.

When it comes to  externally driven reconnection, where oppositely magnetized flows are forced together at an imposed external velocity,   then the question is not how fast  reconnection occurs, but what structures develop to accommodate the imposed merging rate.  That may also  be amenable to laboratory study. 

New laboratory studies are  also being performed  with laser driven 
\cite{Nilson+2006,Willingale+2010,Fox+2012,Fiksel+2014,Rosenberg+2015a,Rosenberg+2015b,Chien+2019} and pulsed power driven \cite{Suttle+2016,Suttle+2018,Suttle+2020,Hare+2017a,Hare+2017b,Hare+2018} experimental platforms. In both types of platforms the plasma flows are supersonic (sonic Mach number $M>5$), for  laser experiments so far they are also super-Alfvenic, while  pulsed power experiments operate at Alfv\'en Mach number 
 $ 0.7\le M_A \le 2$. These experiments demonstrated: pile-up of magnetic flux and formation of shocks at the boundary of the reconnection layer, formation of plasmoids \cite{Hare+2017a} happening at low Lundquist number of ~100 occurring in the semi-collisional reconnection regime \cite{Loureiro+2016,Bhat+2018}, and generation of fast outflows with velocities exceeding the Alfv\'en velocity.

In the context of astrophysical jets however,  the real frontier of reconnection experiments is particle acceleration, for which there has so far been little experimental insight so far, although there may be some hope  \cite{Chien+2019}.  Similarly, although acceleration at collisionless shocks have long been widely studied observationally \cite{Parks+2017}, theoretically, and computationally  e.g. \cite{Marcowith+2016},   experimental studies are  only recently emerging  \cite{Fox+2013,Huntington+2015,Fiuza+2020}.  

Ref. \cite{Fiuza+2020}  shows  that laser plasma shock experiments 
may help to understand how the well-known electron injection problem in shock acceleration e.g. \cite{Treumann+2009} might be overcome.  The injection problem is this:  while ion-mediated shocks are widely invoked phenomenologically  as sites of relativistic electron acceleration, electrons have to be pre-accelerated to reach energies where they can  be further scattered across the shock and participate in  standard diffusive shock acceleration.  These laser driven experiments show that  first-order Fermi-type pre-acceleration of  electrons by Weibel instability-driven turbulence at the shock seems to occur naturally,    offering a solution to the injection problem at least in this particular non-relativistic shock context.

\subsection{MT launch, early collimation, and instability using coaxial gun helicity injection experiments}
\label{coaxial}

Ref. \cite{Bellan2018} provides an excellent  review of 
jet-related laboratory studies with a particular focus on  coaxial gun experiments and so we 
refer the reader to a much more detailed treatment there. 
We give a brief overview  here.

These experiments  consist of two coaxial electrodes
linked in vacuum by an axisymmetric  magnetic field  \cite{Hsu+2002,Hsu+2003,Hsu+2005}.
This is somewhat analogous to an accretion disk linked to central object
via loops of azimuthally distributed poloidal magnetic field.  (see Fig \ref{fig0.1}a-d). At eight azimuthal locations, 
plasma is injected onto to these poloidal fields
and an electric potential is applied between the anchoring
electrodes. This causes
toroidal rotation of the plasma
results that generates a toroidal field  
from  twisting the poloidal field.  In this way, finite magnetic helicity is  injected into the experiment.
For these experiments, the Alfv\'en Mach number is less than unity
and the  launch is  most like the  MT, which it helps 
to elucidate.

Once the magnetic twist is injected and the toroidal field amplified, 
loops rise and merge on the axis,  not unlike what might happen in 
an astrophysical disk.  A twisted unipolar core tower   forms, rises, 
and remains collimated by hoop stress. 
Although it is the unipolar inner core that rises,
the cross section that  would correspond to the physical cross section of the astrophysical MT-type jet 
also includes the downward return magnetic flux.

The amount of twist helicity injected is characterized by the
 free parameter \cite{Hsu+2002,Hsu+2005}
$\lambda_{gun}\equiv {\bf J}\cdot{\bf B}/B^2= \mu_0 I_{gun}/\Psi_{gun}={4\pi\over L q}$,
where $I_{gun}$ is the current from the imposed voltage across
the electrodes, $\Psi_{gun}$ is the initial poloidal
magnetic flux, $L$ is the plasma column length.
Here  $q=2\pi aB_z(a)/LB_\phi(a)$ where $B_z(a)$ is the poloidal field, $B_\phi(a)$
is the toroidal field  and $a$ is the plasma column radius.
The Kruskal-Shafranov condition for  kink instability is $ q <1$ or $\lambda_{gun} > {4\pi\over L q}$ 
in this context \cite{Hsu+2005}.
 
 Measurements   agree with  predictions of the Kruskal-Shafranov kink instability criterion. 
For  $\lambda_{inj}\le 4\pi/L$, 
the collimated structure is stable.  When $\lambda_{gun} \ge 4\pi/L$, the unipolar  core of the tower  forms
exhibits a kink instability, but the structure stays connected.  This is the regime of the 
CCD image sequence shown  in the example of Fig \ref{fig0.1}e.
For $\lambda_{gun}>> 4\pi/L$, the core tower forms,
a kink instability occurs, and a  disconnected  blob 
detaches.   A general message is that the kink instability can occur and   not immediately
destroy the overall  tower core collimation.  This principle  also emerges from  
pulsed-power experiments \cite{Lebedev+2005a} discussed further below .

Unlike the original
MT which was constructed as a series of static force free-equilibria and required
ambient pressure for collimation,  in the coaxial gun experiments, the plasma has finite $\beta \sim 0.02-0.1$ and deviates from   being force free.  Ref. \cite{Bellan2018} explains
how this generalization leads to a self-consistent collimation tendency of the  tower peak, even without ambient
pressure.  The fact that the top of the tower has a poloidal magnetic field turnaround  produces a  
a toroidal current, whose associated Lorentz force slows the MT  propagation near its apex. The pileup of mass further tightens the vertical field, drawing in the toroidal field as well, and further  collimating the tip.
Although the walls of the apparatus  act like an ambient medium, they are far enough from the
region of  the aforementioned collimation that the latter can be considered self-collimated.


These experiments carefully probe aspects of launch region, but  the time scales and dynamical range in length
are still limited compared to what scaled values would be  for real astrophysical jets.  Also the propagation speeds
reached are very low,  and like all laboratory jet experiments so far, none involve anything close to relativistic bulk motion. 
As such, questions such as the  extent to which  MT can propagate many orders of magnitude times the scale of  
 launch region and  remain magnetically dominated, let alone reach relativistic speeds, 
cannot easily be assessed.  Mechanisms of particle acceleration have correspondingly, also  not yet been assessed.


\subsection{Experiments on jet propagation with collimation by conical standing shocks, or  an ambient magnetic wall } 
These experiments provide analogues to  astrophysical  configurations considered  by e.g. Ref. \cite{Canto+1988}.

\subsubsection{Laser ablation of conical targets}
In these experiments \cite{Farley+1999,Shigemori+2000,Kasperczuk+2006,Gregory+2008}, collimated jets are produced by  the radial convergence of ablated plasma and formation of a standing conical shock where  the plasma flow is redirected axially  (Figure \ref{fig1}). These experiments  tested the effects of radiative cooling on  jet collimation by comparing the divergence of the jets formed from different target materials (from CH to Au). They demonstrated that stronger radiative cooling forms more collimated jets. In the early experiments, the jets were formed and propagated in  vacuum, but  this was later modified to include  ambient material \cite{Nicolai+2008,Tikhonchuk+2008}. The density contrast between the jet and the ambient ranged from 0.1 to 10, the higher end of which compares particularly favorably  to 
proto-stellar jets. However, the addition of the ambient material in these experiments also weakened the jet formation, and the jet length was only a factor of few larger than its  diameter. Use of different atomic number gases (He and Ar) allowed  control over the  level of radiative cooling in the ambient plasma and investigation of how this affects the bow shock morphology. 

More recent modification of this experimental approach included  use of more complicated distributions of laser energy deposition on the target. Experiments at the OMEGA laser facility \cite{Gao+2019,Lu+2019} ablated a plastic target using 20 laser beams in a hollow ring configuration. These experiments revealed the presence of self-generated magnetic fields (via the ``Biermann battery" effect) in kinetic energy-dominated jets. The  magnetic field is  generated both in each of the laser produced plasma plumes, and also in the plasma column formed at the radial convergence of the plumes that form a jet. The magnetic field structure in the jet was analyzed using proton radiography,  along with extensive numerical simulations. The collimated magnetized jet in these experiments has a length of $\sim$4mm and 
$\sim$1mm diameter. Simulations of the experiment suggest a sonic Mach number of $\sim$3, plasma beta of $\sim 10$ and Reynolds and magnetic Reynolds numbers $\sim 10^4$. Due to the high temperature of the jet plasma the (thermal) Pecelet number is below unity so thermal conduction is significant. Use of higher-Z dopants in the target could facilitate reaching radiatively cooled regimes.

\subsubsection{Collimation of laser ablated plasma plumes by  externally applied axial (poloidal) magnetic field walls} 

In this class of experiments 
\cite{Ciardi+2013,Albertazzi+2014,Higginson+2017a,Higginson+2017b,Khiar+2019}, formation of  collimated jets with  length  $\sim 20$ times  the initial jet diameter  is achieved using a  strong axial field.
 The  expanding conducting plasma  is re-collimated by magnetically reflection toward the axis (Figure \ref{fig2}). 
 The $B$-field cannot penetrate into the plasma but is compressed and strengthened as the plasma expands.
The corresponding plasma density decrease reduces the  ram pressure below the magnetic pressure. The magnetic "wall"  then redirects the flow toward  the central axis,  forming a  conical shock, once again similar that of Ref. \cite{Canto+1988}. These experiments also exhibit a Rayleigh-Taylor instability  where the flow is redirected by the magnetic field.  When the orientation of the magnetic field was varied with respect to the target 
normal, 
 collimation  is reduced and becomes ineffective at large angles ($\sim 45$deg, \cite{Higginson+2017a}). For a $B$-field  fully  perpendicular to the target normal (so parallel to the target) there is expansion along $B$-field lines, and also the development of magnetic Rayleigh-Taylor instability at the flow-magnetic barrier interface \cite{Khiar+2019}.
  
\subsubsection{Experiments assessing the collimating influence of  ambient density and coaxial wind}. 
\label{5.4.3}
These  purely hydrodynamic experiments  described below mostly employ  targets with  thin foils laser-heated  to high temperatures / pressures. The heated material then expands via a narrow channel into a lower density ambient material.  
One goal  is to look at the properties of bow shock development and at the reverse flow (material reflected from the bow shock boundary) and compare with numerical simulations. 
In this respect, the experiments again focus specifically on the  jet  propagation regime, rather than 
the launch regime. 

Mach numbers are relatively small ($M< 5$) but  Reynolds numbers $\Re\ge 10^6$ are achieved due to high density and small temperature.   Using laser inertial confinement facilities, Ref. \cite{Blue+2005}, using NIF, and \cite{Foster+2005}, using OMEGA for example, carried  out  experiments in which a laser illuminates a thin metallic  target disk, itself placed 
flush against a much thicker washer  with a hole.  The laser ablates the disk
and the resulting plasma exits through the washer hole   as a  collimated supersonic jet that propagates  into  foam.  
 X-ray radiography and back-lighting were used to study its  propagation in the foam.
Ref. \cite{Blue+2005}  studied  the influence of  nozzle 
angle on jet structure  and compared   symmetric (2-D) vs. titled 
(3-D) nozzles.  Turbulence seems to arise earlier  in 3-D than in  2-D, as 
also seen in simulations,  validating the predictions of the 3-D radiative HD code HYDRA 
\cite{Marinak+1996}.  Reynolds numbers in the experiment 
 were $Re\sim 10^7$ but only $Re <10^3$ in the simulations however.


The experiments of Ref. \cite{Foster+2005} on OMEGA 
obtained  Mach numbers up to  $M\sim 5$ 
with  images of turbulent flows, dense plasma  jets, and bow shocks.
 Modeling was carried out with 2-D hyrdo simulations using the RAGE code  \cite{Gittings+2008}. 
 The experiments incurred a  jet-to-foam density ratio of
$\rho_j /\rho_a \sim 1$, which is intermediate between 
YSO  jets which  have $\rho_j > \rho_a$  and AGN jets with $\rho_j < \rho_a$.
However, the latter  are relativistic, and the experiments involve non-relativistic flows.  

A systematic comparison between  low density jets propagating into high density media and high density jets propagating in to low density media remains an opportunity  of interest for comparison with simulations of same.  For example, comparing the structure of two jets with equal   momentum flux   but 
varying  the relative value of density and velocity so that cases with density above and below the 
ambient are studied.   The same could be done comparing jets of equal  energy flux. 


More recent  experiments on the smaller  LULI  facility    have studied (i) nested hydrodynamic flows
 \cite{Yurchak+2014} and (ii) effects of MHD flows  \cite{Albertazzi+2014}. 
The former focused  on the interaction between a central outflow and a surrounding wind. 
As mentioned in section \ref{collimation}, this is  common in astrophysics and of interest for assessing non-magnetic collimation mechanisms.

Ref.  \cite{Yurchak+2014}   studied the interaction between two nested supersonic plasma flows.  To form the  flows, a phase plate was designed to produce a  laser energy distribution with a ($100 \mu$m) central circular spot surrounded by  a thinner ($75 \mu$ m) ring.  Separate targets were made for the spot and ring such that a central  $15 \mu$m thick iron disk  was surrounded by a $15 \mu$m  thick  plastic ring,  both  resting on a CHAl pusher. The laser launched a shock  into   the pusher which  transmitted to the Fe and CH layers. After the shock passed through  the back of the target, supersonic, interacting  plasma outflows were  driven from both the  outer ring and  inner disk. Experimental results compared with  numerical hydrodynamic simulations showed that the outer wind  strongly collimated the inner outflow.  The results   confirm the "shock-focused inertial confinement" mechanism proposed in some  computational astrophysics investigations \cite{Icke+1992,Frank+1996} and demonstrates the basic  concept behind other nested wind scenarios \cite{Blackman+2004}.    Inertial collimation of  of otherwise uncollimated spherical outflows  is also of particular  interest in  the study of    collimated planetary nebula  \citep{Zou+2020} from common envelope evolution.

\subsubsection{Formation of supersonic radiatively cooled jets from ablation plasma flows in conical wire arrays }

In this  pulsed-power approach \cite{Lebedev+2002,Ciardi+2002,Lebedev+2005a,Ampleford+2005,Bocchi+2011,Plouhinec+2014,Valenzuela+2015}
 jet formation  happens via formation of a conical standing shock, similar to the configuration employed in  laser-driven jet experiments. The main difference from the laser driven platforms is a more steady-state nature of the converging flow formation. Plasma is driven for the duration of the current pulse (from $\sim 300$ns to $\sim 2 \ \mu$s), which is longer than the characteristic hydrodynamic times for the material passing through the standing conical shock. Jet formation and ejection reach a quasi-steady state, and the jet length to diameter ratio has exceeded $\sim 30$. Variation of material of the ablating wires (Al, Fe or W) has been used to study  collimating effects of radiative cooling. The dimensionless parameters characterising these laboratory jets are a radiative cooling parameter $\xi$ (ratio of radiative cooling length to jet radius) of  $1\geq \xi \geq 0.01$;  internal Mach number of $5\le M\le20$, Reynolds number of $ Re  \sim 3\times 10^4 $, localisation (mean free path to jet radius $\sim 10^{-4}$).  All are in the parameter range relevant to  YSO jets. 
 
The main results from these experiments are that: (i) increasing radiative cooling increase  jet collimation; (ii) jet formation is robust with respect to azimuthal asymmetries present in the converging conical flow. This includes both small perturbations arising from discrete wire structure, and  significant global deviations from axial symmetry.  
A modification of  the conical wire array set-up, used ``twisted" wire arrays \cite{Lebedev+2005a,Ampleford+2008}.
 In this case, 
an axial component of magnetic field created by the azimuthal component of the current  added and a finite angular momentum to the converging plasma flow. This formed a rotating jet with a hollow density profile. 

Formation of jets in conical wire arrays is dominated by hydrodynamics, but the ablated plasma flowing from the wires does contain a frozen-in toroidal magnetic field, especially in the lower density  plasma halo  surrounding the dense central jet. This field is not  dynamically significant for  jet collimation, but its presence was noticeable.
 in experiments  designed to study jet-ambient interactions. 

\subsection{Pulsed power jet experiments where collimation  depends on combined MHD hoop stress and hydrodynamics}. 

These employ an experimental configuration in which plasma flows are created by ablation of thin foils driven by a radially-directed, MA-level currents in pulsed-power devices \cite{Suzuki-Vidal+2009,Ciardi+2009,Gourdain+2010,Suzuki-Vidal+2012,Gourdain+2013,Byvank+2016,Byvank+2017}. Here the current rapidly heats the foil, allowing ablation from the top surface. 
A significant rise in the foil resistivity  allows partial penetration of toroidal magnetic field into the ablated plasma. Thus a ${\bf J}\times {\bf B}$ force acting on the ablated plasma accelerates it axially  to high velocity. The absence of foil ablation in the foil centre above the electrode leads to inward plasma motion 
enhanced by  hoop stress from the toroidal magnetic field, which  also drives  a standing conical shock. The overall outflow consists dense  highly collimated jet, surrounded by a lower density magnetized plasma with toroidal field moving with the jet velocity  (Figure  \ref{fig3}).  

What collimates this jet?   The  jet opening angle ($<3$deg in these experiments) is much smaller than what  could be caused simply by the sonic Mach number 2-3, as  calculated using the measured jet temperature.  The temperature and density of the plasma surrounding the central jet are also too small for thermal pressure to be the cause of confinement.  The measured magnetic field at the jet boundary is also too small to balance the jet thermal pressure. Instead, the pressure needed for collimation is provided by the ram pressure of the converging plasma surrounding the jet, driven by the magnetic hoop stress at larger radii.  So it is ultimately magnetically collimated, but non-locally so. 

\subsection{Experiments on Jet interactions with ambient clouds and cross-winds}. 
Here we concentrate on experiments that  separate the jet-ambient interaction from the jet formation, mainly
 on  experiments performed with pulsed-power driven jets. 
 
\subsubsection{Deflection of a jet ejected from a conical wire array by a plasma cross-wind }
\cite{Lebedev+2004,Lebedev+2005b,Ampleford+2007,Ciardi+2008}. A transverse plasma flow (a cross-wind) propagating across the jet path was produced by ablation of a plastic foil from XUV radiation emitted from the standing conical shock and the ablating wires (Figure \ref{fig4}). The main lessons  from these experiments
are (i) the  jet was not destroyed and maintained good collimation even after a $\sim$30 deg change in direction;  (ii) variation of the jet/wind density ratio also leads to variation in the deflection angle and the bending trajectory of the laboratory jet is well described by an astrophysical model \cite{Canto+1995}. (iii) The  experiments on jet-wind interactions were well modeled by simulations using the same numerical code  that simulated astrophysical jet bending with  the same dimensionless parameters  such as Mach numbers and the density and velocity ratios \cite{Ciardi+2008}. In addition to reproducing the overall bending trajectory, the simulations showed  that the jet-wind interaction correctly  predicted the formation of clumps, or variability in the density and emission intensity in the jet. This  happened in a configuration where both the jet and  wind were initially uniform and steady, and the perturbations resulted from the combination of Kelvin-Helmholtz and Rayleigh-Taylor instabilities at the jet-wind boundary. Simulations also looked at the effects of jet rotation on the features formed in the interaction, predicting  observable differences compared to a non-rotating jet.

\subsubsection{Jet interaction with a stationary gaseous cloud} 

\cite{Ampleford+2005,Suzuki-Vidal+2013a}. In these experiments,  jets ejected from a conical wire array or a radial foil, propagated into  a vacuum before interacting with a gas cloud with a relatively sharp boundary. The interaction formed a broad bow shock and an enhanced emission region at the head of the jet. The motion of the working surface in the cloud was in good agreement with  hydrodynamic pressure balance considerations for the measured density contrast between the jet and the cloud, as used to describe internal shocks in astrophysical jets. In addition, experiments showed that the shape of the bow shock was determined not only by the reverse flow of the plasma from the decelerated head of the jet, but also by the presence of a lower density plasma halo  surrounding the central jet and moving with the same velocity. Interestingly, the brightest region of the interaction region  drifted laterally. This was interpreted as  due the presence of an advected toroidal magnetic field in the halo plasma and  asymmetry in return current path, causing an unbalanced ${\bf J}\times {\bf B}$ force  that displaced the jet.

A similar platform  to the experiments of \cite{Foster+2005} discussed in section \ref{5.4.3}, 
was also used to investigate interaction of supersonic jets experiencing glancing collisions with a dense object in conditions scalable to astrophysical YSO jets interacting with molecular clouds \cite{Hartigan+2009}. These authors investigated how the morphology of the deflected bow-shock {Figure \ref{figHart}) depends on the interaction geometry and   numerically simulated the experiment extensively. The experiment reveals the formation  filamentary structures in the working surface of the deflected bow shock, and entertainment of  obstacle material into the flow.      

\subsection{Experiments studying  internal shocks from jet velocity variability}
Observed variation of astrophysical jet emission along the jets is commonly attributed internal shocks arising from ejection variability at the source, whereby faster flow then catches up to slower flow.
In a reference frame moving with the internal shock velocity, the interaction is equivalent to an interaction of two counter-propagating plasma flows, which is a more practical frame in which  to perform laboratory experiments.  Such interactions were produced and studied in \cite{Suzuki-Vidal+2015,Valenzuela+2015} using pulsed-power platforms. In \cite{Suzuki-Vidal+2015} the counter-streaming jets were formed by two ablating radial foils, driven by the same current pulse (Figure \ref{fig5}). The two interacting jets, from the top and bottom foils respectively, move with the same velocities but have slightly different diameters and densities. This difference comes from  two-fluid MHD effects (reversal of current direction), found to be important in the jet-launching regions of radial foil jets \cite{Gourdain+2013}. The jet propagating from the top foil has a smaller diameter and a higher density and  ram pressure, and the interaction with the second jet results in the formation of a bow shock slowly moving downwards (see Fig.2 of \cite{Suzuki-Vidal+2015}. The overall motion of the bow shock is consistent with  standard shock models. The formed bow shock is initially smooth, but then develops small-scale structures that lead to  clumps  and  fragmentation.  The spatial and temporal scales of this process are consistent with the thermal instabilities developing at the appropriate slope of the radiative cooling curve.  

Ref.  \cite{Suzuki-Vidal+2015} includes discussion of the scaling of the observed instabilities to the conditions of typical Herbig-Haro objects,  concluding that these would correspond to 3-30AU clumps evolving on a $\sim 15$ yr time-scale. Formation of clumps  was also observed in experiments with colliding plasma jets produced by two conical wire arrays \cite{Valenzuela+2015}.                

\subsection{Magnetically dominated jets using pulsed power machines }

In contrast to coaxial gun experiments which include the arguably more realistic  combination of
 toroidal and poloidal fields in the MT  \cite{Bellan2018}, 
  pulsed-power driven experimental configurations have mostly used a purely toroidal magnetic field to drive the outflow. There have however been attempts to add a weak axial magnetic field (see item 3 below).
  
 Experiments discussed below are most related to MT jet models and involve formation of a magnetically dominated cavity with a central, magnetically confined jet, evolving inside a overall ambient  plasma.
The experiments  used a radial wire array or radial foil configuration made of two concentric electrodes connected radially by thin wires or a thin foil \cite{Lebedev+2005a, Lebedev+2005b,Ciardi+2007,Suzuki-Vidal+2010}. When driven by a $\sim  1$MA, $\sim 300$ns current pulse, this configuration produces an axially directed flow of ablated plasma which forms the surrounding plasma environment for the MT formation. This stage starts when the wires (or foil) near the central electrode fully ablate 
 triggering formation of a magnetic bubble and a magnetically dominated jet (Figure \ref{fig7}).  

The main results from these experiments are:
{\it 1. Formation of a magnetically dominated cavity confined by an ambient plasma}. The ambient plasma can be generated by the ablation flow from the radial wire array or radial foil, in which case the ambient plasma moves upwards with velocity comparable to the  axial expansion of the magnetic cavity.  Alternatively, the ambient density can be initially introduced as a stationary gas cloud above the radial foil configuration. The evolution of the cavity is determined by the current flowing through the central jet and the walls of the cavity. The pressure of the toroidal magnetic field associated with this current, drives both the radial and axial expansion of the cavity, and the axial acceleration of the central jet. The magnetic cavity expands much faster axially  than  radially,   as determined by the magnetic pressure inside the cavity and the density of the ambient plasma ahead of the cavity.  The axial expansion velocity of the cavity equals the Alfv\'en velocity calculated with the magnetic field inside the cavity and the ambient density ahead of it.  The  cavity interior is magnetically dominated, with little plasma. 

{\it 2. Formation of a dense current-currying jet along the axis of the magnetic cavity.} This jet 
is confined by the toroidal magnetic field and  rapidly becomes MHD unstable, with the development of $m=0$ and $m=1$ modes. The growth times for these modes are consistent with MHD predictions and are much shorter ($\sim$2 ns) than the cavity evolution time ($\sim 200$ ns), but the experiments show that the instability does not completely disrupt the jet. Instead, the outflow becomes clumpy, with strong density variations, while remaining well-collimated (Figure \ref{fig8}). The level of instability is partly determined by the balance between the cavity evolution time and the time required to short-circuit the current at the base of the jet and thus reducing the current remaining in the central jet. Strong radiative cooling in the central jet    plays a significant role in the observed \ collimation of the outflow. The observed morphology of the jet evolution  agrees with numerical  modeling using both laboratory plasma and astrophysical codes \cite{Ciardi+2007,Huarte-Espinosa+2012}  - see Fig. 20, 21 from \cite{Lebedev+2019}. Current-driven instabilities of the central jet column were also experimentally observed and computationally simulated in \cite{Gourdain+2012} for magnetic tower jets formed in radial foil configurations.  

{\it 3. Influence of axial magnetic field}. This was investigated in \cite{Suzuki-Vidal+2010}. In these experiments the jets were formed using the same radial wire arrays as before, but an initially weak axial magnetic field was added using a coiled return current conductor or a coil inserted into the cathode electrode. The high magnetic Reynolds number of  this system allowed compression of the field by the radially converging plasma  to values comparable with the toroidal magnetic field in the jet. The axial field increased the jet diameter and decreased the x-ray emission, indicating a smaller plasma temperature.  The axial magnetic field was insufficient to stabilize the MHD instabilities which were still observed, and the formation of clumps was not affected. Overall, the presence of the axial magnetic field in the central jet at the level comparable with the toroidal magnetic field did not have a dramatic effect on the jet launching in these experiments. Effects of axial magnetic field on the magnetic tower jets formed in radial foil configuration were also experimentally and computationally studied in \cite{Byvank+2016, Byvank+2017}.       

{\it Diffusive acceleration of energetic ions in a MT} Magnetic cavities in experiments with MT jets have strong toroidal magnetic fields capable of confining high energy ions. The injection of ions (predominantly protons) into the cavity may occur due to the voltage applied to the base of the cavity or during clump formation. Measurements with a magnetic spectrometer and  proton imaging techniques  showed that the proton energy spectrum extended from $\sim$100keV to $\sim$3MeV, and that the protons originate from the clumpy jet \cite{Suzuki-Vidal+2013b}.  The maximum energy of the energetic protons exceeded the maximum voltage used to drive the experiment by at least an order of magnitude and the proton Larmor radius with this energy was comparable to the cavity size (``Hillas constraint", \cite{Hillas+1984}). The observed generation of energetic particles was interpreted in \cite{Lebedev+2019} as due to diffusive acceleration from random perturbations of the magnetic field formed from  MHD instabilities in the central jet.     

{\it Episodic MT  jet formation} In the MT jet model,  formation of the jet is a transient process, and experiments and  previous radial wire array configurations investigated properties of  outflows formed in a single episode of magnetic cavity formation. However, closure of the radial gap through which the magnetic flux is injected into the cavity by  plasma  restores the initial current path, allowing subsequent episodes of  jet re-formation. This was achieved using a radial foil configuration with appropriately adjusted diameter of the central electrode and of the foil thickness \cite{Ciardi+2009,Suzuki-Vidal+2009,Gourdain+2010}. A succession of multiple magnetic cavities and embedded jets were observed to propagate over length scales spanning more than order of magnitude (Figure \ref{fig9}). The central jets in each of the cavities were unstable to  current driven instabilities developing on  Alfv\'en time-scales. A second important time-scale is the cavity ejection period, which was a factor of 10-20 longer and determined by the temporal variability of the Poynting flux injection at the base of the system. The resulting outflow is heterogeneous and clumpy, and propagates along a well-collimated channel made of nested cavities. Formation of the episodic magnetically dominated jets in these experiments also accompanied by episodic burst of soft x-ray generated during compression of the central jet.   

\subsection{Rotating plasmas in high energy density experiments}
Though laboratory experiments are unable to imitate rotation in a gravity field of a central object, they can form radially converging plasma flows with non-zero angular momentum for limited studies of sheared flows and transformation of radially convergent flows into axial jets \cite{Ryutov+2011}. Stagnation of such flows near the axis leads to formation of a rotating disc or cylinder which is supported in equilibrium by the ram pressure of the flow. In pulsed-power driven experiments, rotating plasma flows were created using ablation plasma flows in cylindrical wire arrays which were modified to add an angular momentum to the flow. This was achieved by either using a cusp configuration of axial $B$-field to add a radial component magnetic field at the wire positions, or by using an azimuthal displacement of return conductors positioned close to the ablating wires \cite{Bocchi+2013a, Bocchi+2013b,Bennett+2015}. The main observations from these experiments can be summarised as follows:  plasma convergence toward the axis of the system led to the formation of a hollow, rotating plasma disc characterised by a high Reynolds number $(Re \sim 10^5)$. Supersonic rotation of this disc  sustained at a constant value of the outer radius for few rotation periods, due to the ram pressure of the converging flow. At the same time,  the inner boundary of the disc experienced a rapid inward expansion on a time-scale  4 to 5 orders of magnitude shorter than the classical plasma diffusion time. The large Reynolds number characterising this system and the large initial perturbations in the flow from the discrete nature of the plasma streams forming the disc, could allow for the development of turbulent viscosity on a time-scale comparable with one rotation period. The observed fast inward motion would imply an effective turbulent viscosity of five orders of magnitude exceeding the classical plasma viscosity (see discussion in \cite{Lebedev+2019}).

\section{Conclusion}

Many persistent  questions remain to be answered in the pursuit of  understanding  jet physics in each of the three regimes of launch, propagation, and particle acceleration.  Having first  summarized  some of these long standing issues,  we then discussed  laboratory astrophysical  experiments performed so far that pertain to these  themes.  
These  included brief discussions of  dynamos, magneto-rotational instability, magnetic reconnection and particle acceleration, experiments, followed by more extensive discussion of   jet formation and propagation experiments.  Our review complements  others  \cite{Bellan2018,Lebedev+2019} both in the breadth of topics discussed and in that  previous reviews typically begin with the experiments  and  discuss applications. In our specific  discussion of  jet experiments, we also  in our  focus more on pulsed power and laser driven experiments rather than coaxial gun experiments covered in Ref. \cite{Bellan2018}, although we have discussed all of the above. 

Laboratory experiments  do offer the substantial benefit of a controlled laboratory environment,  but also pose substantial challenges such as required high powers, short time scales, and small lengths scales needed for practical apparatus to achieve regimes of relevance.  As we have seen,  the enterprise has included numerous  creative efforts and  results 
  that provide proof of principle, and even new insight on  specific plausible physics components  of jet engine,  formation,   propagation, and dissipation physics. 
However,  the approach that we have taken in this review---starting with the open questions first--reveals substantial gaps  between what has been accomplished to date and  what could be called a direct solution of an open astrophysical question.  There is  an  opportunity for the future to close this gap, but it may require a different way of thinking about these laboratory experiments.    

Perhaps the fact that most existing experiments
 employ facilities not designed specifically to answer  astrophysical questions  is a  limitation that warrants  substantial effort to overcome.    Mapping the approach used for  this review onto a strategy for future work, we might advocate thinking about  what new apparatus could be built to address specific astrophysical jet questions,  rather than  creatively having to retrofit experimental apparatus designed originally for very different purposes. Summarized succinctly, this  means shifting  from the  more common approach of  "What can we do with  existing laboratory facilities  that might have astrophysical application?"  to  "What facilities should we build in the future to best answer a specific astrophysical question?" 
 
In the absence of specialised astrophysics-inspired lab facilities, it is necessary to continue to utilize  all possible opportunities at  existing HEDP facilities to perform well scaled and well diagnosed experiments to address relevant astrophysical issues. It is especially important to use these  opportunities to test numerical simulation tools in astrophysics with real experimental data, since benchmarking of astrophysics codes on real observational data remains rather limited.

\section{Acknowledgments}
EB acknowledges 
support from NSF Grants  AST-1813298, PHY-2020249,  DOE grant DE-SC0020432, DE-SC0020103, KITP UC Santa Barbara funded by  NSF Grant PHY-1748958, and Aspen Center for Physics funded by NSF Grant PHY-1607611. SL acknowledges support from DE-NA0003764, DE-F03-02NA00057, DE-SC-0020434, and DE-SC-0001063 by US AFOSR grant FA9550-17-1-0036, and by the Engineering and Physical Sciences Research Council (EPSRC) Grant No. EP/N013379/1


\bibliographystyle{elsarticle-num}
\bibliography{blackmanlebedev}

 \begin{figure}[t!]
\includegraphics[width=0.7\textwidth]{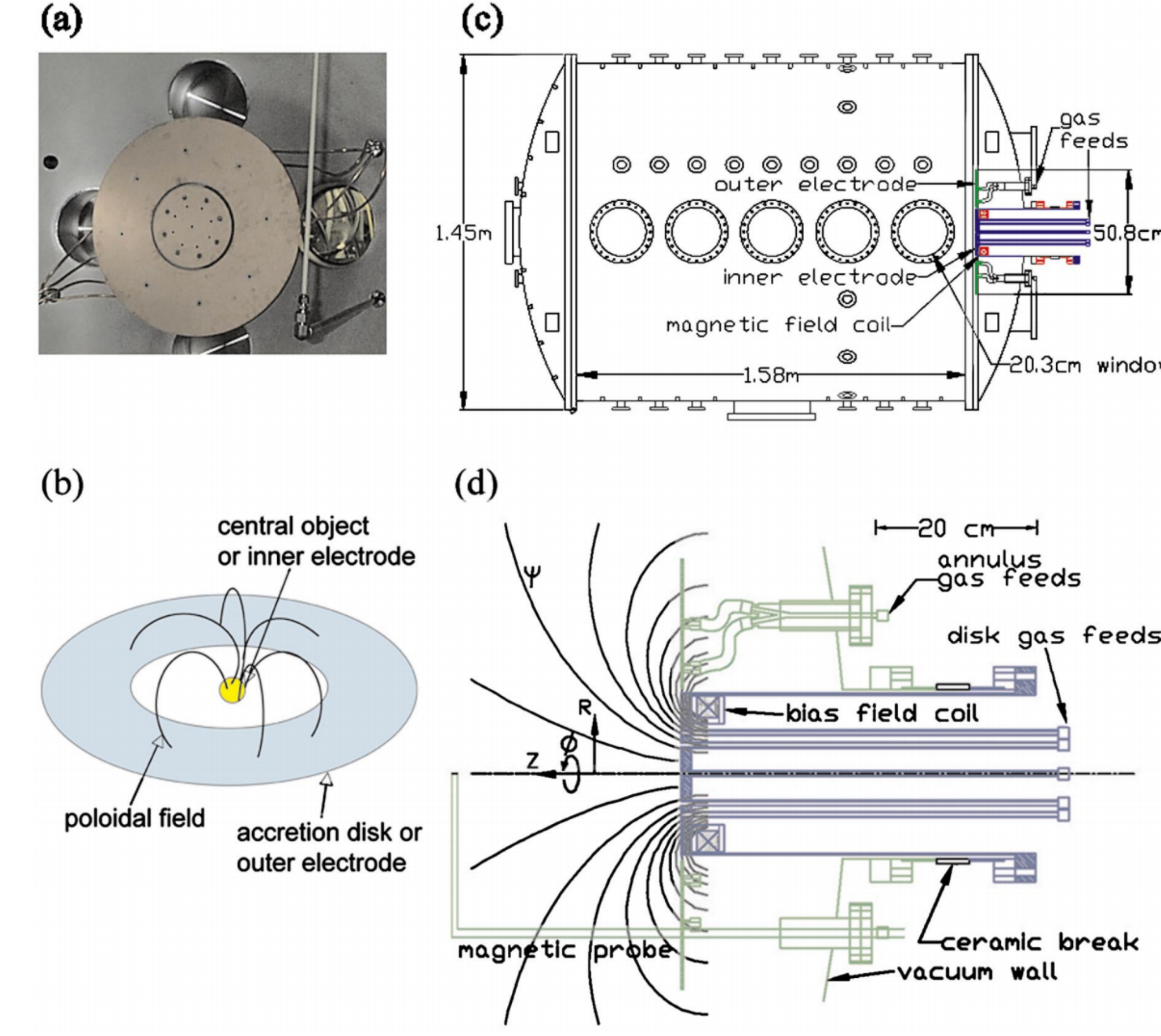}
\vskip-0.8in
\includegraphics[width=0.6\textwidth,angle=-90]{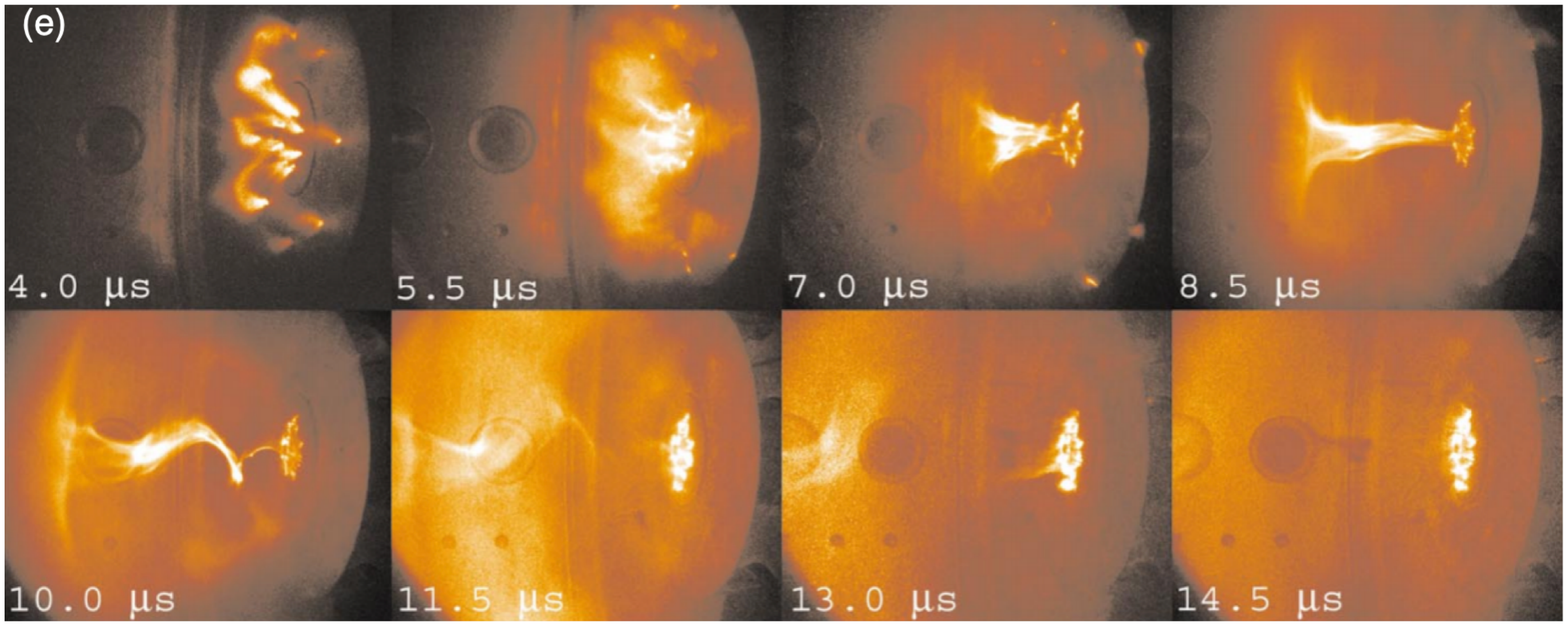}
\caption{Combined  from figs 1 and 3b of Ref.  \cite{Hsu+2005}: (a): Coaxial gun and the rotatable magnetic probe array.
(b): Schematic showing analog between the  coaxial gun and
astrophysical disk. (c):Experimental vacuum chamber with  diagnostic ports and coaxial gun are mounted at the right end.  (d):Side-view schematic the  coaxial gun, showing inner electrode (blue), outer electrode (green), gas feed lines, contours of constant initial poloidal flux, and cylindrical coordinates. (e). CCD images of plasma evolution for typical intermediate $40 {\rm m}^{-1}\leq \lambda_{gun}\equiv \mu_0 I_{gun}/\Psi_{gun}\le 60 {\rm m}^{-1}$, a marginally unstable regime in
which the central column becomes helical  but does not immediately detach from base. 
system.}
\label{fig0.1}
\end{figure}

 \begin{figure}[t!]
\includegraphics[width=0.7\textwidth]{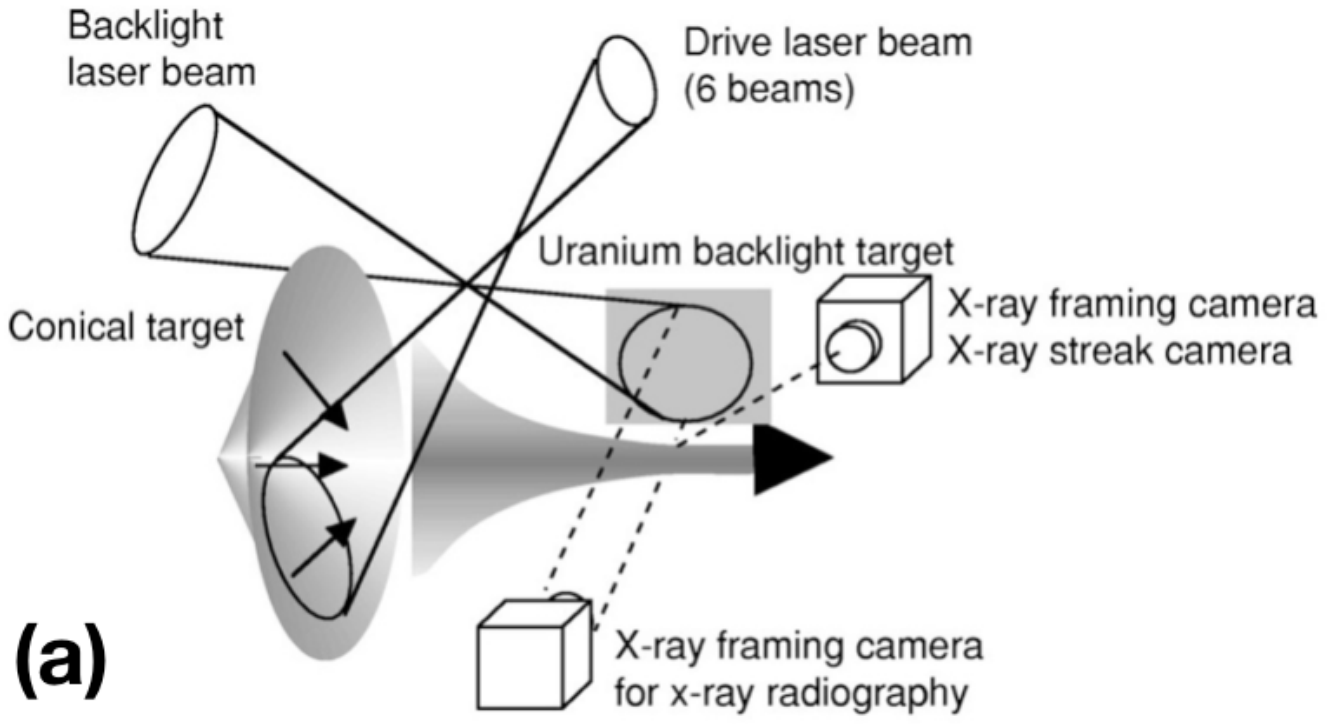}
\hskip-1.5in
\includegraphics[width=0.7\textwidth]{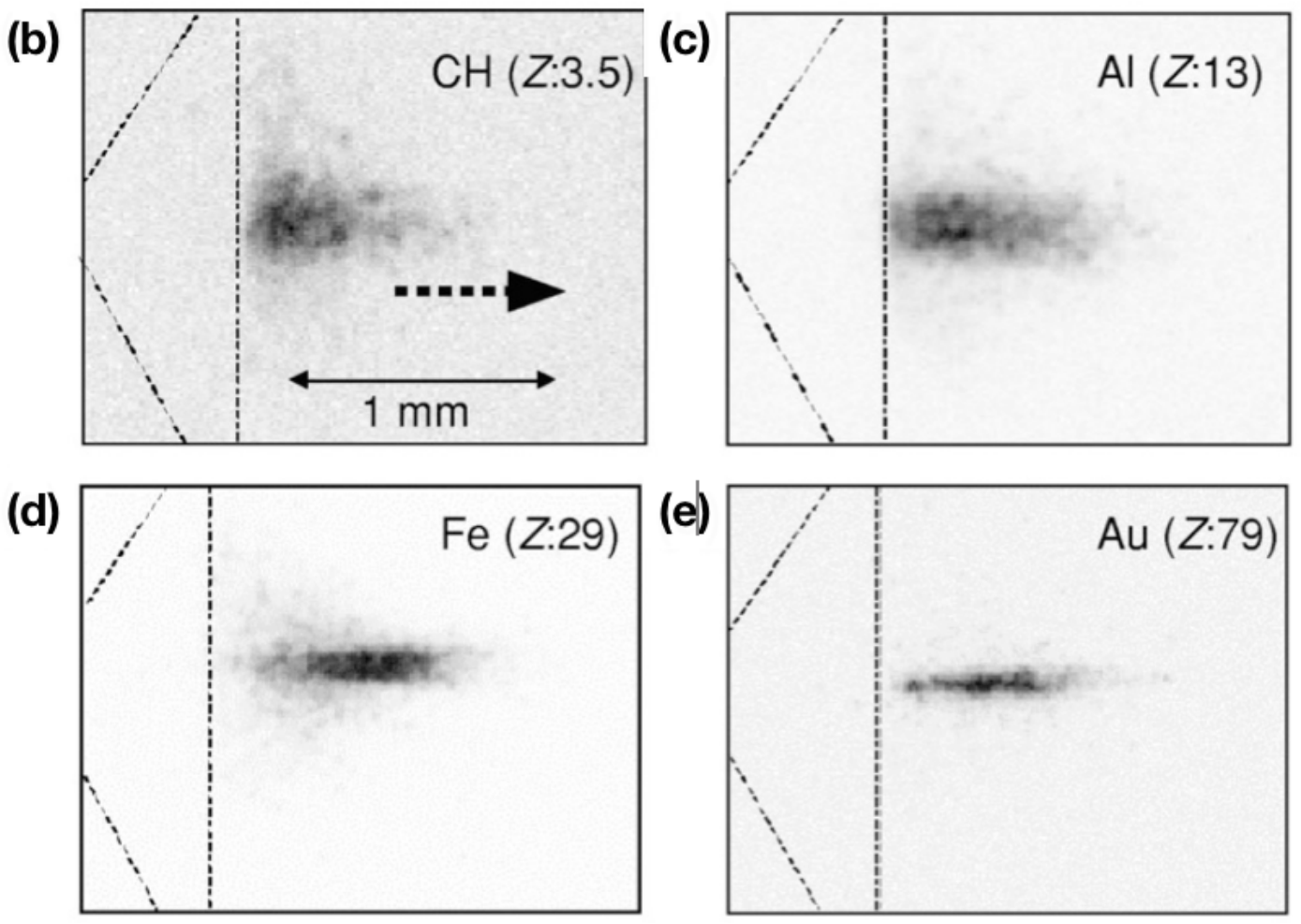}
\vskip-1.5in
\caption{From Ref.  \cite{Shigemori+2000}: (a): Schematic of the laser ablation driven radiative jet experiment
(b-e):  Self-emission images of laser-produced jets for different levels of radiative cooling.}

\label{fig1}
\end{figure}

 \begin{figure}[t!]
\includegraphics[width=0.9\textwidth]{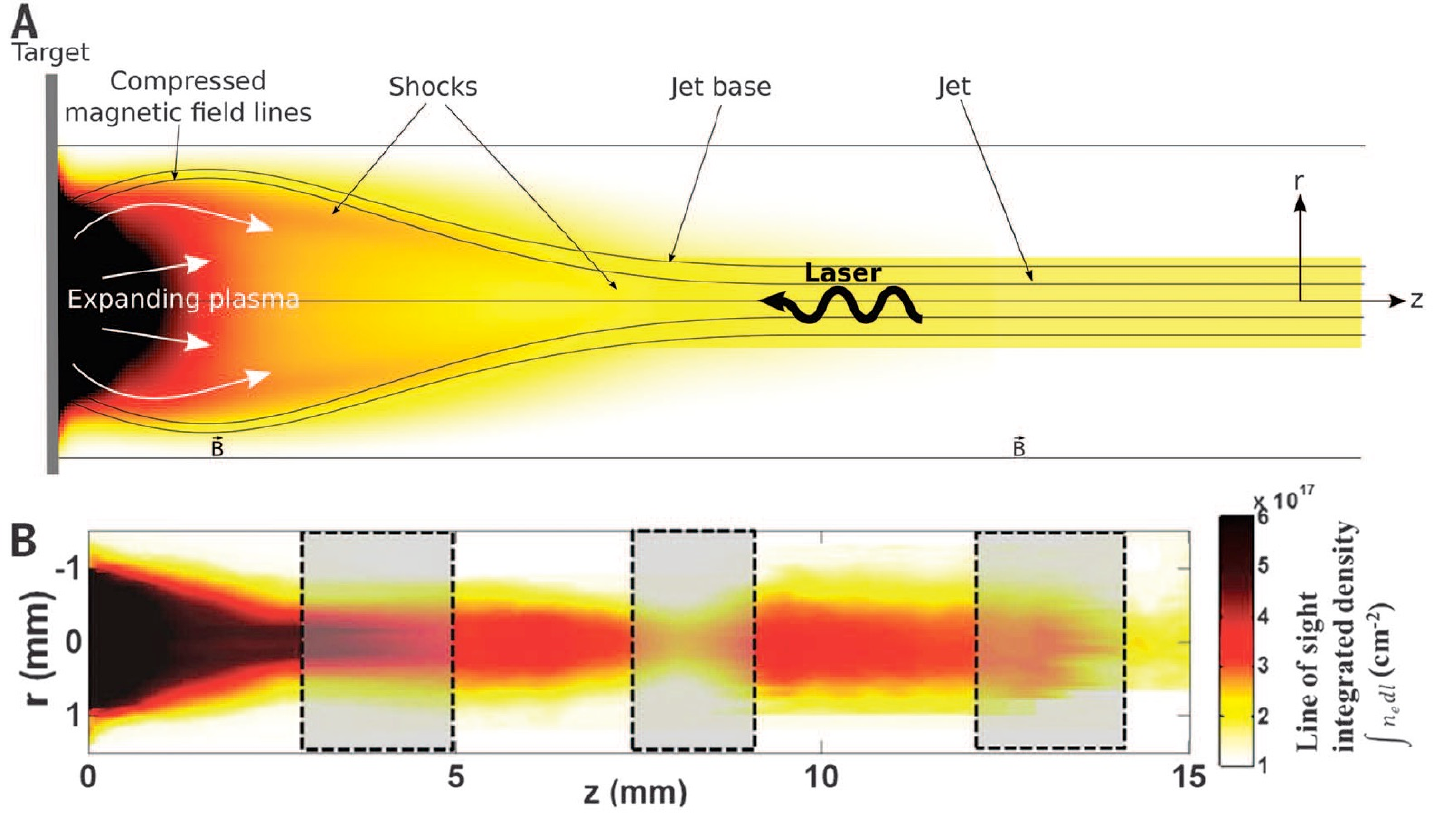}
\caption{(A) Schematic of  pulsed-power experiment demonstrating  jet-formation that depends on collimation from  both hydrodynamic focusing and MHD hoop stress, overlayed on plasma line density image. (B) Composite image of the plasma line density measured with interferometry \cite{Albertazzi+2014} and plasma density legend.}
\label{fig2}
\end{figure}

 \begin{figure}[t!]
\includegraphics[width=0.9\textwidth]{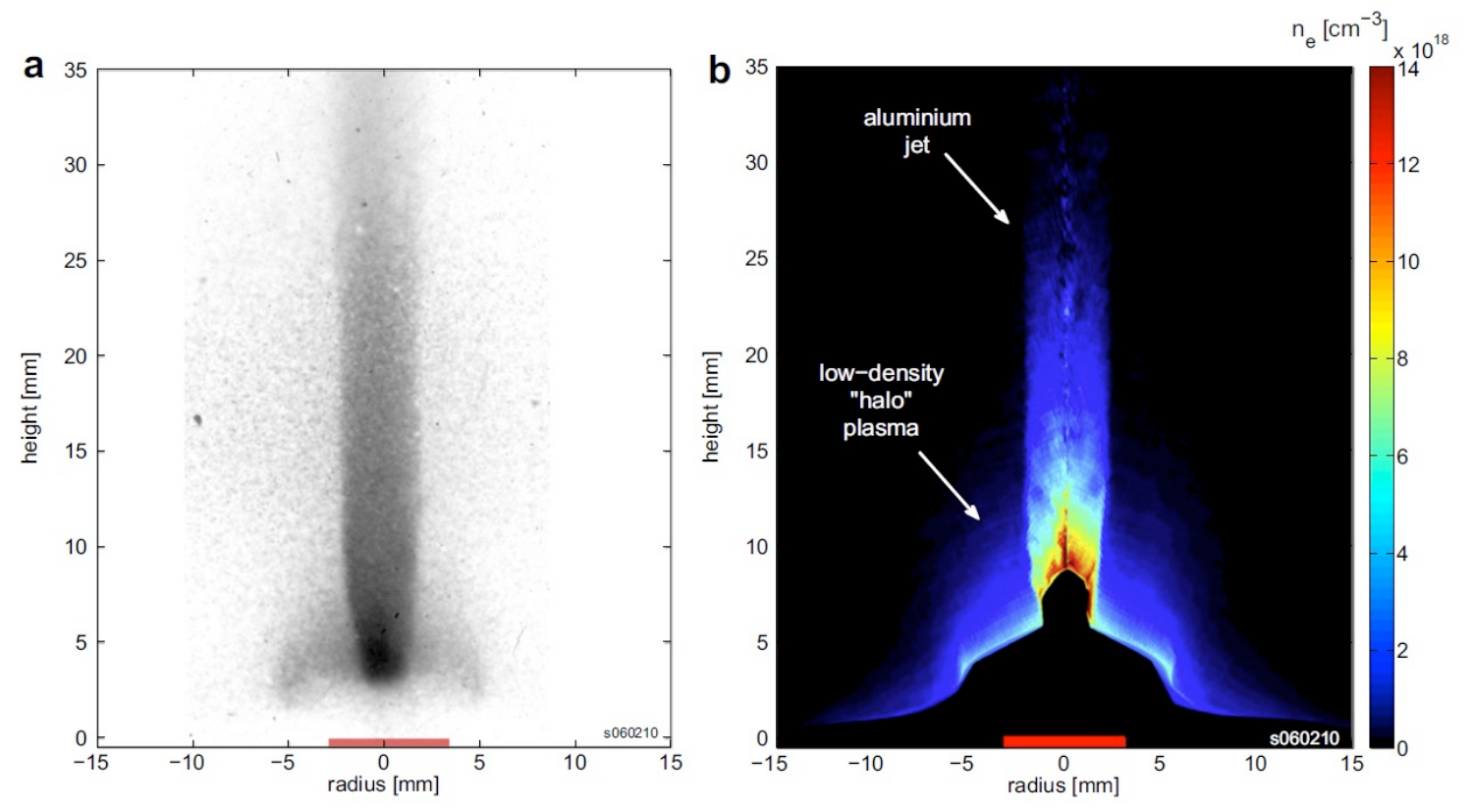}
\caption{Jet formed in radial foil set-up. (a) XUV self-emission; (b) density map measured by laser interferometry \cite{Suzuki-Vidal+2013a}.}
\label{fig3}
\end{figure}

 \begin{figure}[t!]
\includegraphics[width=0.9\textwidth]{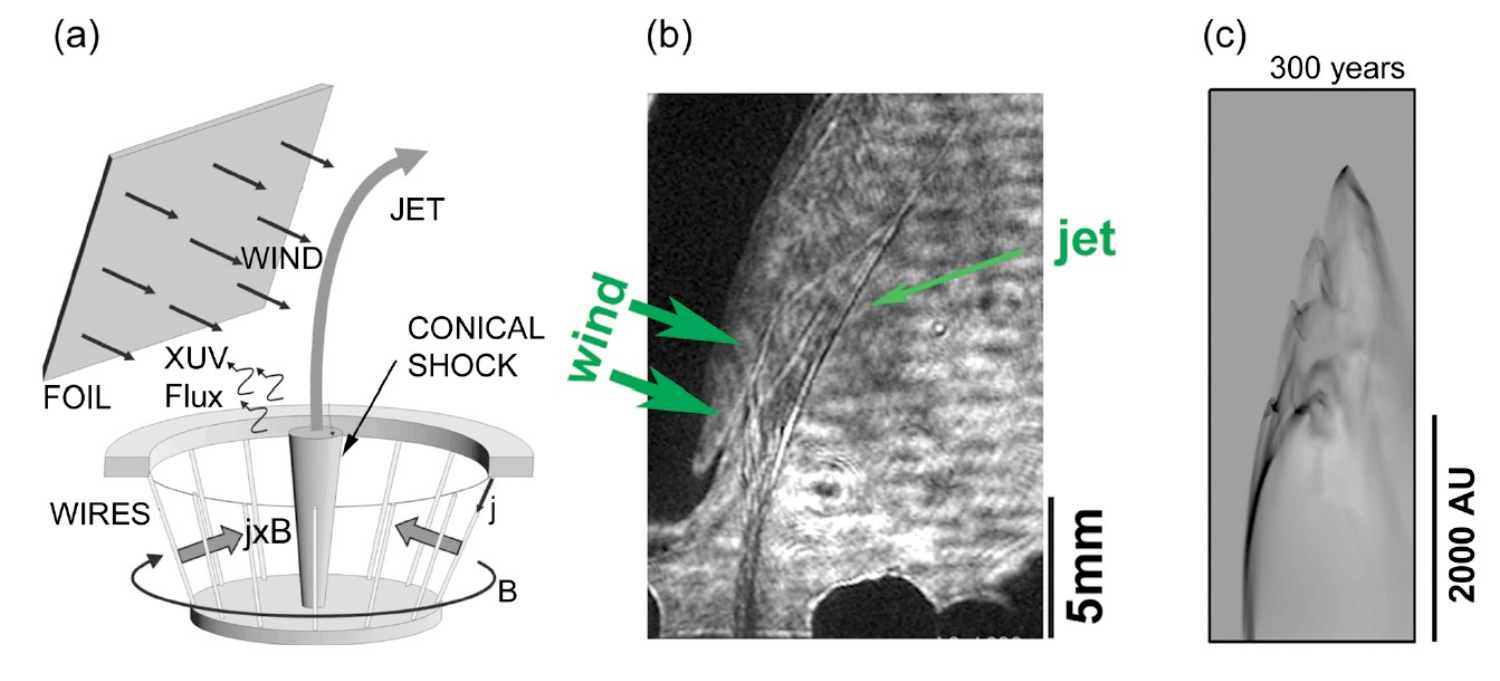}
\caption{(a): Schematic of the jet formation in a conical wire array and a side-wind generation by surface plasma expansion. (b): time-resolved laser Schlieren image of jet bending. (c): density structure of an astrophysical jet interacting with a cross-wind at similar dimensionless parameters (Adapted from \cite{Ciardi+2008})}.
\label{fig4}
\end{figure}

 \begin{figure}[t!]
\includegraphics[width=0.9\textwidth]{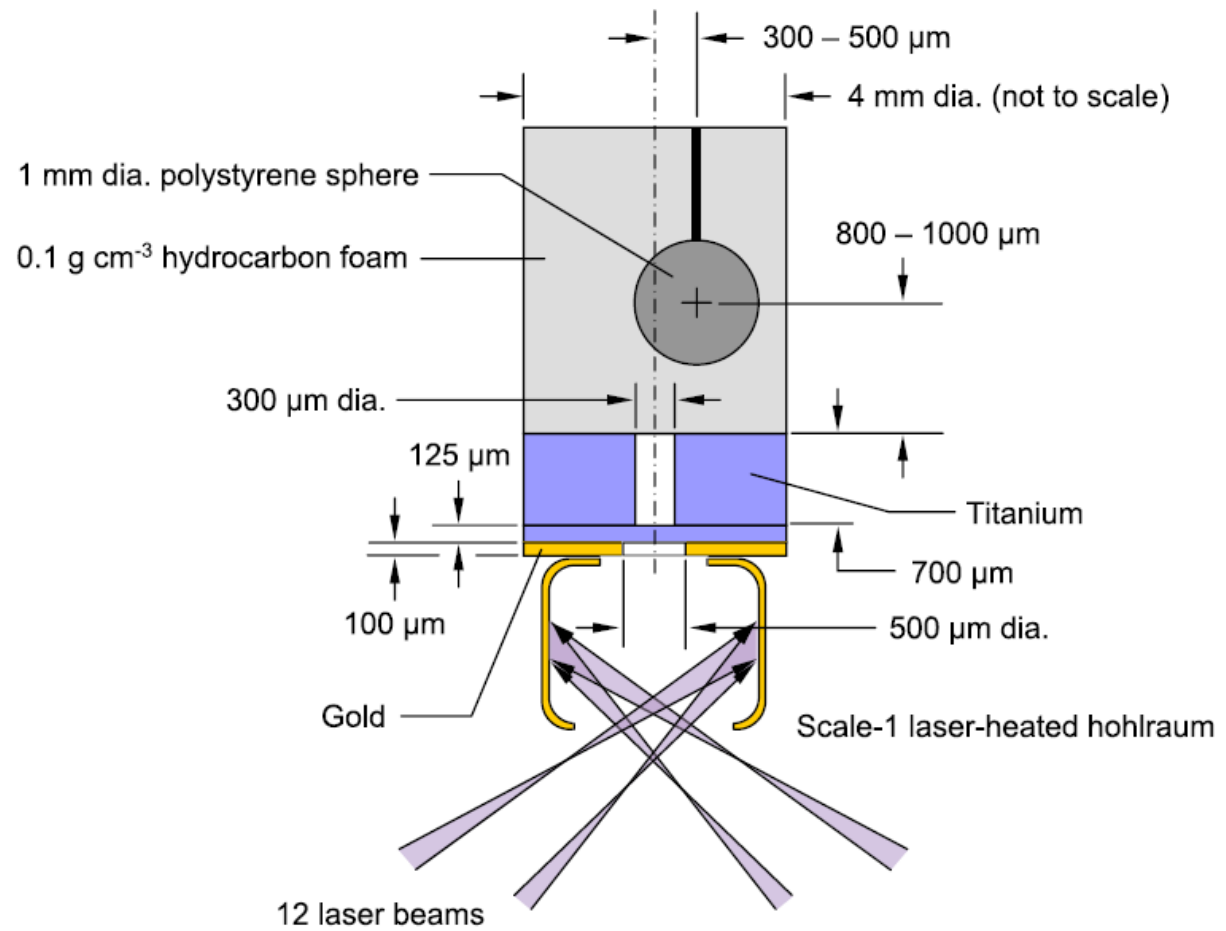}
\includegraphics[width=0.9\textwidth]{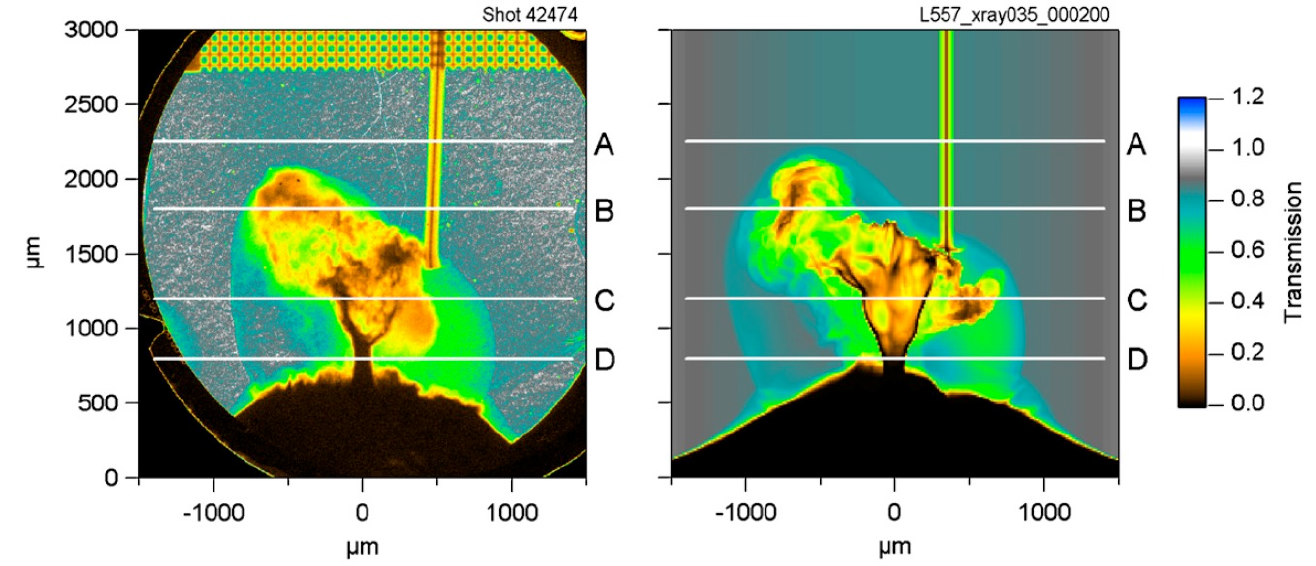}
\caption{Jet deflection experiments of Ref. \cite{Hartigan+2009}. (a)
Sketch of the experimental set-up for jet deflection experiments on Omega laser facility .(b) Observed (left) and simulated (right) X-radiography images of jet deflection by the obstacle }.
\label{figHart}
\end{figure}

 \begin{figure}[t!]
\includegraphics[width=0.7\textwidth]{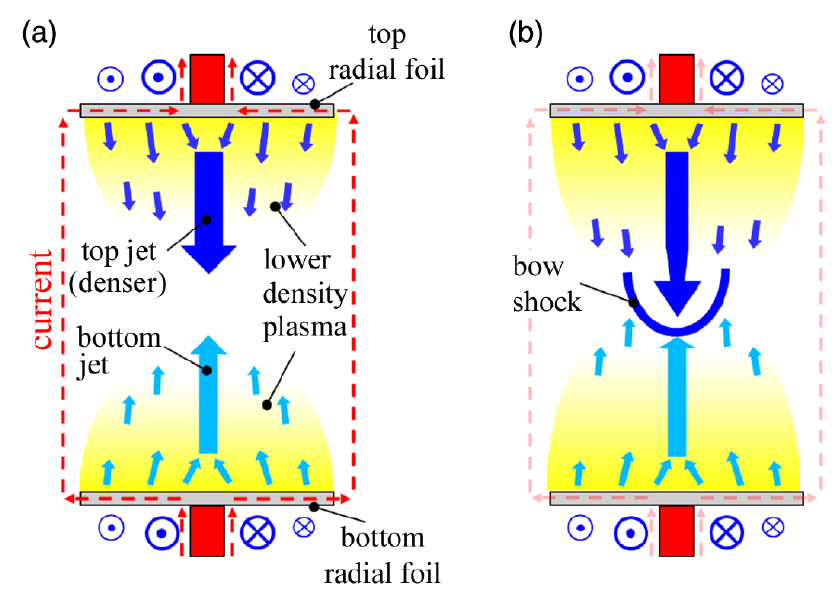}
\caption{Schematic of experimental configuration to study the formation of a bow shock from the interaction of counter-streaming jets with different diameters and ram pressures \cite{Suzuki-Vidal+2015}.}
\label{fig5}
\end{figure}

 \begin{figure}[t!]
\includegraphics[width=0.7\textwidth]{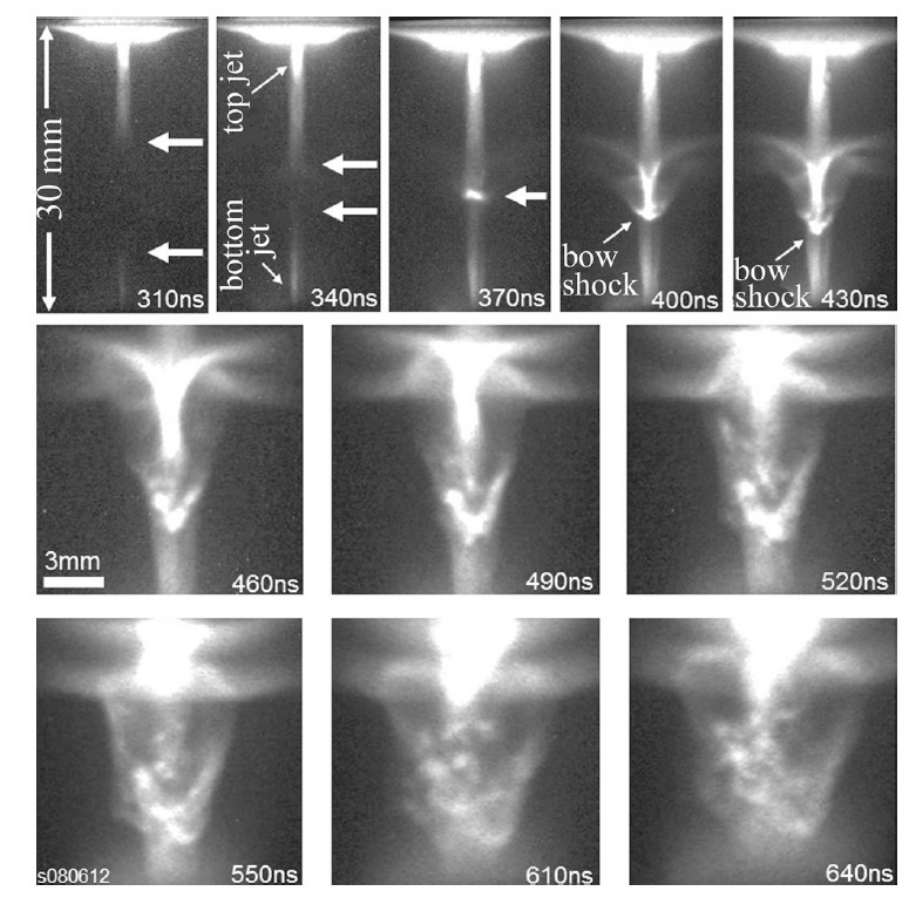}
\caption{ Self-emission optical images from the same experiment as Fig. \ref{fig5} showing the formation of a bow shock and its fragmentation from radiative cooling instabilities \cite{Suzuki-Vidal+2015}.}
\label{fig6}
\end{figure}

 \begin{figure}[t!]
\includegraphics[width=0.7\textwidth]{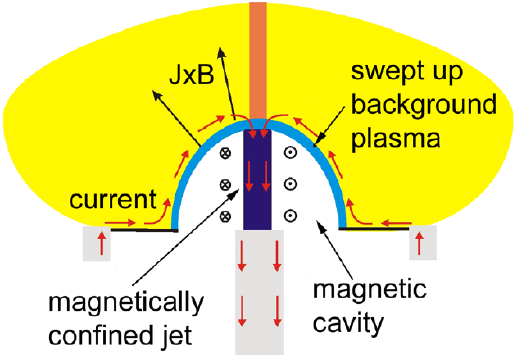}
\caption{ Schematic of magnetic tower jet formation in radial wire array experimental set-up \cite{Lebedev+2005a}.}
\label{fig7}
\end{figure}

 \begin{figure}[t!]
\includegraphics[width=0.7\textwidth]{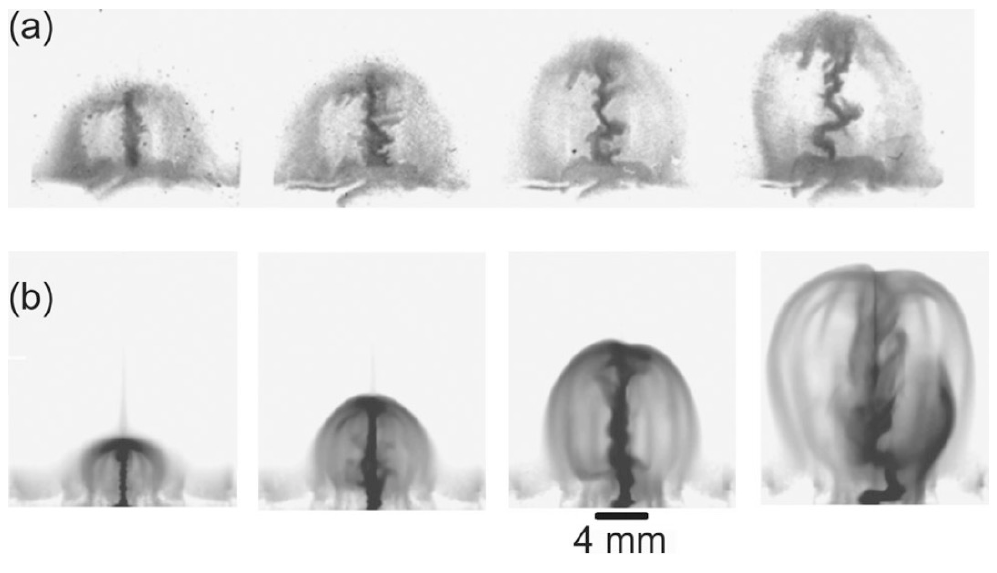}
\caption{ (a) Temporal evolution of magnetic tower jet in experiment (XUV self-emission) and (b) in MHD simulations of the experiment \cite{Ciardi+2007}.}
\label{fig8}
\end{figure}

\begin{figure}[t!]
\includegraphics[width=0.9\textwidth]{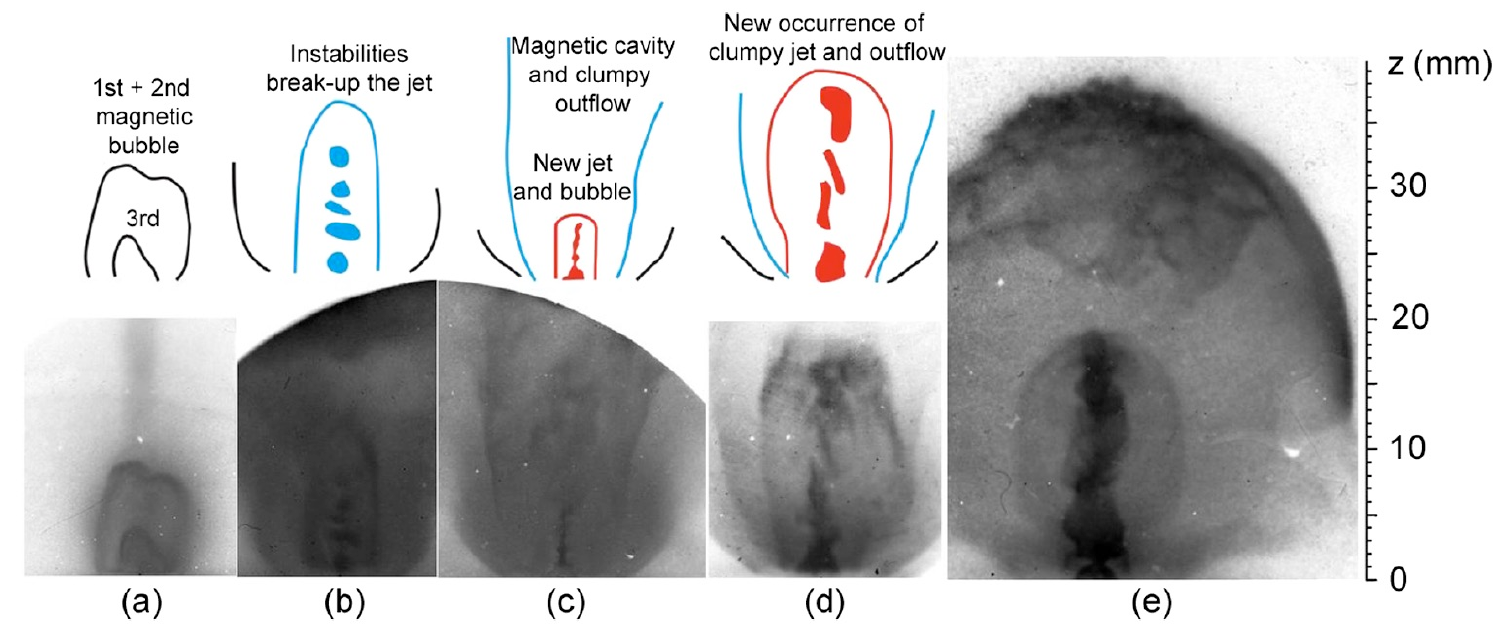}
\caption{ Time sequence of soft x-ray images showing the evolution of episodic magnetic tower jet (from 
\cite{Ciardi+2009}.}
\label{fig9}
\end{figure}

\end{document}